\documentclass[twocolumn,aps,prx,footinbib,floatfix,nolongbibliography,superscriptaddress]{revtex4-2}
\usepackage[utf8]{inputenc}

\usepackage{amsmath,amssymb,bm,ulem}
\usepackage{xcolor,graphicx,siunitx}
\usepackage[colorlinks=true, linkcolor=blue, citecolor=blue, pdfencoding=auto]{hyperref}
\usepackage{tikzsymbols}
\usepackage{slashed}

\newcommand{\ket}[1]{|#1\rangle}

\begin{document}

\newcommand{\thetitle}{Topologically protected flatness in chiral moir\'e heterostructures}
\title{\thetitle}

\author{Valentin Cr\'epel}
\thanks{These two authors contributed equally}
\affiliation{Center for Computational Quantum Physics, Flatiron Institute, New York, NY 10010, USA}
\author{Peize Ding}
\thanks{These two authors contributed equally}
\affiliation{Department of Physics, Columbia University, New York, NY 10027, USA}
\author{Nishchhal Verma}
\affiliation{Department of Physics, Columbia University, New York, NY 10027, USA}
\author{Nicolas Regnault}
\affiliation{Laboratoire de Physique de l’\'Ecole normale sup\'erieure, ENS, Universit\'e PSL, CNRS, Sorbonne Universit\'e, Universit\'e Paris-Diderot, Sorbonne Paris Cit\'e, 75005 Paris, France}\affiliation{Department of Physics, Princeton University, Princeton, NJ 08544, USA}
\author{Raquel Queiroz}
\affiliation{Center for Computational Quantum Physics, Flatiron Institute, New York, NY 10010, USA}
\affiliation{Department of Physics, Columbia University, New York, NY 10027, USA}

\begin{abstract}
The observation of delicate correlated phases in twisted heterostructures of graphene and transition metal dichalcogenides suggests that moiré flat bands are intrinsically resilient against certain types of disorder.
Here, we investigate the robustness of moir\'e flat bands in the chiral limit of the Bistrizer-MacDonald model -- applicable to both platforms in certain limits -- and demonstrate drastic differences between the first magic angle and higher magic angles in response to chiral symmetric disorder that arise, for instance, from lattice relaxation. 
We understand these differences using a hidden constant of motion that permits the decomposition of the non-abelian gauge field induced by interlayer tunnelings into two decoupled abelian ones. 
At all magic angles, the resulting effective magnetic field splits into an anomalous contribution and a fluctuating part. 
The anomalous field maps the moir\'e flat bands onto a zeroth Dirac Landau level, whose flatness withstands any chiral symmetric perturbation such as non-uniform magnetic fields due to a topological index theorem -- thereby underscoring a topological mechanism for band flatness. 
Only the first magic angle can fully harness this topological protection due to its weak fluctuating magnetic field. 
In higher magic angles, the amplitude of fluctuations largely exceeds the anomalous contribution, which we find results in a physically meaningless chiral operator, and an extremely large sensitivity to microscopic details and an exponential collapse of the single particle gap. 
Through numerical simulations, we further study various types of disorder and identify the scattering processes that are enhanced or suppressed in the chiral limit. Interestingly, we find that the topological suppression of disorder broadening persists away from the chiral limit and is further accentuated by isolating a single sublattice polarized flat band in energy. Our analysis suggests the Berry curvature hotspot at the top of the $K$ and $K'$ valence band in the transition metal dichalcogenide monolayers is essential for the stability of its moir\'e flat bands and their correlated states.
\end{abstract}

\maketitle

\section{Introduction}

The recent observation of the fractional quantum anomalous Hall effect in twisted monolayers of $\rm MoTe_2$ marks the first unequivocal evidence of topologically ordered states in quantum material without external magnetic fields~\cite{cai2023signatures,xu2023observation,zeng2023thermodynamic,park2023observation}.
It paves the way for realizing non-abelian states of matter~\cite{mong2014universal,vaezi2014superconducting,repellin2018numerical,barkeshli2016charge,barkeshli2011bilayer,barkeshli2012topological,crepel2019matrix,katzir2020superconducting,reddy2024non,crepel2024attractive,Cr_pel_2024}, which are essential for the development of topologically protected quantum computing~\cite{kitaev2003fault}. 
However, topologically ordered states are typically fragile to perturbations. 
The original fractional quantum Hall effect in $\rm GaAs$ was only observed for extremely high mobility samples~\cite{stormer1999fractional}.
In comparison, transition metal dichalcogenides (TMDs), such as $\rm MoTe_2$, are not necessarily clean monolayers. 
Their mobility usually lies two orders of magnitude below that of typical samples of $\rm GaAs$ that host fractional quantum Hall states~\cite{movva2015high,chuang2016low,xu2017odd,pisoni2019absence,zhao2023fractional}. 
Additionally, strain, lattice relaxation, and twist angle variations are expected to further disorder the heterostructures built from these monolayers~\cite{Wilson2020,carr2019exact, Nakatsuji2022,sainz2021high,thomson2021recovery}. 
Nonetheless, fractional Chern insulators~\cite{sheng2011fractional,neupert2011fractional,regnault2011fractional,Cr_pel_2023} were observed in both twisted $\rm MoTe_2$ homobilayers and twisted bilayer graphene (TBG) at finite field~\cite{xie2021fractional}.
This unexpected observation hints at the possibility that these heterostructures carry some intrinsic protection to a disorder that enables the realization of such delicate, strongly correlated phases.

In this work, we ask the question: Can the flatness of moir\'e flat bands be itself topologically protected? 
A hint towards an answer follows from the pioneering work by Aharonov and Casher~\cite{aharonov1979ground}, who showed that a relativistic Dirac particle in a magnetic field has exactly $\Phi/\Phi_0$ zero modes of definite chirality, where $\Phi$ is the total magnetic flux through the lattice, and $\Phi_0$ the flux quantum. 
As a consequence, the zeroth Landau level of a Dirac particle remains perfectly flat even when the magnetic field is non-uniform. 
This perfect immunity against certain forms of disorder is tied to a chiral anomaly and a topological index that protects the net chirality of the system~\cite{katsnelson2007graphene}.
The anticommuting chiral symmetry relates states of opposite energies, allowing the assignment of a proper chirality to all zero energy states, such as the zeroth Dirac Landau level. 
These states with definite chirality must remain strictly at zero energy, even in the presence of disorder, as long as the chiral symmetry is preserved. 
The same anomaly guarantees that the previous theory cannot be realized on a lattice since the total chirality of fermions on a lattice must vanish~\cite{nielsen1981no}. 
Therefore, a chiral Landau level must only appear as a surface theory~\cite{zhang2012surface} or within a valley isolated in momentum space. 
The latter scenario is relevant for the $K$ valley in both graphene~\cite{novoselov2005two} and TMDs~\cite{xiao2012coupled}.
The chiral anomaly has remarkable consequences when it comes to the effect of disorder in the Landau levels of these materials: 
Provided the disorder does not couple the two Dirac valleys of opposite chirality, their zeroth Landau level is not broadened by disorder, unlike every other Landau level at a finite energy~\cite{kawarabayashi2009quantum}. 
A similar result is expected for Landau levels of quadratic bands, provided their $g-$factor is exactly two resulting from the Berry curvature at the band edge~\cite{dubrovin1980ground}. 
No anomaly or chiral Landau levels are expected for a gas of free electrons under a magnetic field, such as in $\rm GaAs$.

With an external magnetic field, which is typically homogeneous, the additional protection due to the chiral anomaly has limited consequences. 
Only a slight broadening difference is seen in experiment between the zeroth and higher Landau levels in graphene~\cite{giesbers2007quantum}. 
In fact, in Hall systems, protection by a chiral anomaly has little physical importance because the Landau level spread to cyclotron gap ratio can be efficiently reduced by using larger magnetic fields. 
In contrast, moiré materials do not rely on an external magnetic field. Instead, they feature a fictitious gauge field arising from non-uniform interlayer couplings. 
The same couplings are responsible for the spectral gaps around the band of interest, which cannot be arbitrarily tuned using external fields. 
As a result, the protection offered by the index theorem can play a crucial role.

In this paper, we explore the protection of flat bands in moir\'e bilayers endowed from the chiral anomaly in the chiral limit of the Bistrizer-MacDonald model of TBG~\cite{bistritzer2011moire,tarnopolsky2019origin}, which has also been shown to be an insightful limit for twisted TMDs upon adding a mass to the Dirac theory of each monolayer~\cite{crepel2023chiral}, and may also apply to other types of heterostructures~\cite{khalaf2019magic,guerci2022higher,wan2023topological,wang2022hierarchy,crepel2023chiralbis, labastida2023flat, crepel2024efficient}. 
This limit permits a formal map between the flat bands at the magic angle and Landau levels under a pseudo, non-abelian gauge field $\vec{\mathcal{A}}$. 
Naively, we should \emph{not} expect protection by the index theorem, which only applies when the gauge field is abelian (Sec.~\ref{sec_indextheorem}). 
Yet, we discover a hidden non-local symmetry that allows for the decomposition of the $SU(2)$ non-abelian gauge field into two decoupled abelian components $U(1)\times U(1)$ (Sec.~\ref{sec_theory}). 
In the decoupled basis, electrons feel a non-uniform and periodic magnetic field $B'$. 
The matching of the magnetic length introduced by this effective magnetic field and the moir\'e length provides an analytical criterion for magic angles, as anticipated in Ref.~\cite{parhizkar2023generic} where it was understood as a consequence of the topological Atiyah-Singer index theorem~\cite{atiyah1968index,atiyah1984dirac}.
For all but the largest magic angle however, the fluctuating part of the magnetic field is pathological, and the system cannot fully harness the topological protection offered by the index theorem for reasons we detail in Sec.~\ref{sec_disorder}. 
First, the index theorem only protects the system with respect to chiral symmetric disorder. 
The large and short-scale oscillations of the effective magnetic field obtained at the higher magic angles corresponds to a quickly oscillating chiral operator, and, in turn, to a very fine-tuned set of disorder realizations keeping the flat bands perfectly flat. 
In contrast, the chiral operator at the first magic angle is mostly carried on low harmonics of the moir\'e lattice, with which realistic disorder pattern have high overlap, yielding an effective protection of the band flatness. 
Second, fluctuations of the effective magnetic field $B'$ in higher magic angle are much greater than its average, which is known to yield an exponential squeezing of the spectrum around zero energy~\cite{becker2022mathematics,hormander1960differential,duistermaat1973global,zworski2001remark}. 
As a result, non-chiral disorder will affect higher magic angles more strongly than the first. 
Altogether, this leads to striking qualitative differences in the robustness of the different magic angles against chiral preserving disorder that we numerically investigate by introducing random higher harmonics of the moir\'e interlayer tunneling potential. 
The first magic angle is mostly undisturbed by higher harmonics, which makes it possible to consistently define the same magic angle condition throughout the sample despite small deformations.
This is not true for higher magic angles, suggesting that it is more difficult to observe fragile correlated phases in higher magic angles in realistic experimental settings. 
In Sec.~\ref{sec_tmd}, we consider the case of twisted TMDs where the anomaly applies thanks to the Berry curvature hotspots at the top of the $K$ and $K'$ valence band.

\section{Index theorem} \label{sec_indextheorem}

A highly counter-intuitive property of the zero-energy Landau level of Dirac particles in a magnetic field was simultaneously discovered by Aharonov and Casher~\cite{aharonov1979ground} and Dubrovin and Novikov (who considered the square of the Dirac Hamiltonian)~\cite{dubrovin1980ground}: it remains \textit{perfectly flat} and extensively degenerate for \textit{arbitrary non-uniform} magnetic fields provided the total magnetic flux threading the system is a constant. 
This behavior strikingly contrasts with that of all higher Landau levels of the same system and those stemming from quadratic bands, which in similar settings acquire a dispersion proportional to the magnetic field's non-uniformity. 

The perfect immunity of the zeroth Landau level to magnetic field variations has since then been interpreted as a consequence of a topological index theorem~\cite{katsnelson2007graphene} formulated by Atiyah and Singer (AS) for Dirac operators coupled to arbitrary (multi-dimensional) gauge fields in even dimensions~\cite{atiyah1968index,atiyah1984dirac}. 
The two-dimensional version with matrix valued gauge field $\vec{\mathcal{A}} = (\mathcal{A}_x, \mathcal{A}_y)$ relevant for the present study states that the operator
\begin{equation} \label{eq_genericdirac}
\mathcal{H}_{\rm D} = \begin{bmatrix} 0 & \mathcal{D}^\dagger \\ \mathcal{D} & 0 \end{bmatrix} , \quad \mathcal{D} = - 2 i \mu_0 \partial_{\bar{z}} + \mathcal{A}_z , 
\end{equation}
with $2 \partial_{\bar{z}} = \partial_x + i \partial_y$, $\mathcal{A}_z = \mathcal{A}_x + i \mathcal{A}_y$ and $\mu_0$ the identity with the same size as $\mathcal{A}_{z}$, is subject to the topological anomaly
\begin{equation} \label{eq_genericindex}
{\rm dim} ~ ( {\rm Ker } ~ \mathcal{D} ) - {\rm dim} ~ ({\rm Ker } ~ \mathcal{D}^\dagger ) =  \int_{S} \frac{{\rm Tr} \mathcal{B}}{2\pi} , 
\end{equation}
where $S$ is the compact surface on which $\mathcal{H}_{\rm D}$ is defined, and $\mathcal{B} = \vec{\nabla} \times \vec{\mathcal{A}} + [\mathcal{A}_x ,  \mathcal{A}_y ]$ denotes the curvature corresponding to the gauge field, \textit{i.e.} the magnetic field strength. 
The block structure in Eq.~\ref{eq_genericdirac} is supplied by the anticommuting chiral symmetry that the Dirac Hamiltonian enjoys in even dimensions. 

The topological nature of the AS theorem is clear: it only depends on the integrated magnetic flux but not on the local details and variation of $\mathcal{B}$. 
The right-hand side of Eq.~\ref{eq_genericindex} is nothing but the Chern number of the electromagnetic connection, or gauge field, with which the reader may be more familiar in the context of Berry connection/curvature~\cite{thouless1982quantized} (to avoid confusion, we stress that these two Chern numbers are \textit{a priori} unrelated and that the ``topology'' of the AS theorem is distinct from to band topology). 
The most direct physical consequence of the AS theorem is that the number of zero modes of $\mathcal{H}_{\rm D}$ is \textit{independent of} local variations of the gauge potential $\vec{\mathcal{A}}$ and of the metric of the Dirac operator. 
Put differently, the density of states at zero energy is topologically protected against all perturbations of the Hamiltonian anticommuting with the chiral operator. 
In the case of graphene in a magnetic field, these perturbations physically correspond to strain and lattice corrugations~\cite{kawarabayashi2009quantum,katsnelson2008zero,phong2023protected}.

The AS index also protects the zeroth Landau level of systems with quadratic dispersion provided they can be written down as the square of $\mathcal{H}_{\rm D}$ up to an overall sublattice dependent uniform potential~\cite{dubrovin1980ground,estienne2023ideal}. This condition is equivalent to having a $g$-factor precisely equal to 2. In some contexts, it can lead to spin-orbit-like corrections to the quadratic Hamiltonian~\cite{crepel2023chiral}.

While the index theorem enforces the existence of an extensive number of zero modes, there is a cost to pay for this flatness when fluctuations of the magnetic field are large compared to its average: the spectral gap separating the zero modes from the rest of the spectrum become exponentially small in the amplitude of magnetic field fluctuations. 
This exponential squeezing of the full spectrum as a function of the maximum amplitude of the effective gauge field $\mathcal{A}$ was proven for the specific case of the chiral Hamiltonian of TBG in Ref.~\cite{becker2022mathematics}, and follows from very generic considerations on Landau operators~\cite{hormander1960differential,duistermaat1973global,zworski2001remark}. 
Intuitively, this scaling of the gap comes from the fact that the existence of zero modes requires a discrete spectrum of the Hamiltonian around which the resolvent is large~\cite{dencker2004pseudospectra}, allowing to form eigenvectors with energy $\delta E$ away from the zero modes for perturbations of the Hamiltonian that are exponentially small in $\delta E$. 

In realistic settings, disorder does not solely occur in the chiral channel. It nevertheless carries a chiral component, against which the system is protected, and an non-chiral one. When the chiral symmetry varies slowly on the moir\'e scale, the harmless chiral part will be large as the most important contribution of the disorder carried by the low moir\'e harmonics strongly overlap with the chiral protected channels. This is what happens at the first magic angle. 
In contrast, when the chiral operator wildly fluctuates within the moir\'e unit cell, the overlap of any realistic disorder realization will have no overlap with it and be almost entirely unprotected. This is what we observe at the higher magic angle, which we discuss in detail in Sec.~\ref{sec_theory} and \ref{sec_disorder}.

\section{Application to chiral TBG} \label{sec_theory}

In this section, we introduce the chiral model for TBG and prove that it decouples into two independent Dirac Hamiltonians akin to Eq.~\ref{eq_genericdirac} coupled to abelian gauge fields. 
In this abelian basis, flat bands appear when the total magnetic field flux per moiré unit cell equals $2\pi$ due to the AS theorem, which provides an intuitive picture for the emergence of the magic angles.
The magnetic field distribution in the unit cell at the different magic angles will be crucial to understanding the robustness of the flat bands against disorder, investigated in Sec.~\ref{sec_disorder}.

\subsection{Interlayer tunneling as a fictitious gauge field}

We consider the chiral limit of the Bistritzer-MacDonald model for TBG~\cite{bistritzer2011moire,tarnopolsky2019origin}, which in the basis $\{(A,t),(A,b),(B,t),(B,b)\}$, with $A/B$ and $t/b$ the sublattice and layer degrees of freedom respectively described by $\vec{\sigma}$ and $\vec{\mu}$ Pauli matrices, and in valley-$K$ reads
\begin{align} \label{eq_chirltbg}
\frac{\mathcal{H}}{|v \kappa|} & = \begin{bmatrix} 0 & \mathcal{D}^\dagger \\ \mathcal{D} & 0 \end{bmatrix} ,  \quad \mathcal{D} = - 2 i \mu_0 \partial_{\bar{z}} + \alpha \mathcal{A}_z ,  \\
\mathcal{A}_z & = A_z^{(1)} \mu_1 + A_z^{(2)} \mu_2 , \quad  \alpha = \frac{w}{|v \kappa|} , \notag \\
A_z^{(1)} & = \!\! \sum_{n=0,1,2} \!\! \omega^n \cos ( \kappa_n \cdot r ) , \quad A_z^{(2)} = \!\! \sum_{n=0,1,2} \!\! \omega^n \sin ( \kappa_n \cdot r ) , \notag
\end{align}
with $\kappa_n$ the $2\pi n/3$-rotation of the $\kappa$ corner of the moir\'e Brillouin zone aligned with the negative $y$-axis, $|\kappa| = 2 |K| \sin (\theta/2)$, $v$ the Fermi velocity of graphene, and where $\alpha$ measures the strength of $AB$ tunneling $w$ in unit of the typical moir\'e kinetic energy  $|v\kappa|$. 
The Hamiltonian in the other valley is obtained by time-reversal conjugation. The system has $C_{3z}$, $C_{2x}$, $C_{2y}T$ and $C_{2z}T$ symmetries, with $T$ the time-reversal operator~\cite{song2021twisted}.

The suggestive form of Eq.~\ref{eq_chirltbg} makes it clear that $AB$ interlayer tunnelings in chiral TBG act as a fictitious gauge potential for the two coupled Dirac cones with a chiral anticommuting symmetry $\sigma_3$~\cite{liu2019pseudo}. 
Furthermore, the chiral model Eq.~\ref{eq_chirltbg} is known to host perfectly flat bands at zero energy for $\alpha = 0.586, 2.221, 3.751, \cdots$~\cite{becker2022mathematics,becker2021spectral}. 
In that context, it is natural to wonder whether these exact flat bands are topologically protected by the index theorem of Eq.~\ref{eq_genericindex} or not~\cite{sheffer2021chiral,parhizkar2023generic}.
At first sight, the answer to this question seems to be negative. 
Indeed, $\mathcal{A}_z$ physically comes from inter-layer tunneling and is entirely carried by off-diagonal layer Pauli matrices. 
As a result of the trace in Eq.~\ref{eq_genericindex}, the index corresponding to this gauge field must vanish. In fact, this is a generic feature of SU($N$) gauge fields with $N>1$, whose generators are all traceless.

\subsection{Hidden constant of motion} \label{ssec_hiddenconstantmotion}

We overcome this problem by proving that an invertible transformation can decouple the Hamiltonian of chiral TBG (Eq.~\ref{eq_chirltbg}) into two independent abelian $U(1)$ Dirac operators, for which the AS index theorem may apply. Our goal is now to find an invertible transformation $U$ making the Dirac operator abelian or diagonal; in other words, we want 
\begin{equation} \label{eq_transformeddirac}
\mathcal{D}' = U^{-1} \mathcal{D} U = -2i\mu_0 \partial_{\bar{z}} + A_z' \mu_3 , 
\end{equation}
for some $A_z'$. When such a $U$ exists, we notice that $[\mu_3, \mathcal{D}'] = 0$, such that the original Dirac operator possesses a hidden constant of motion $[U \mu_3 U^{-1}, \mathcal{D}] = 0$. Conversely, if $\mathcal{D}$ has a constant of motion of the form $\vec{X} \cdot \vec{\mu}$, then the matrix $U = N^{-1/2} (\vec{X}\cdot \vec{\mu} + x \mu_3)$, with $N = 2 x (x + X_3)$ and $x^2 = \vec{X}\cdot \vec{X}$, provides the desired invertible operator for Eq.~\ref{eq_transformeddirac} to hold. Referring to App.~\ref{app_abelianization} for a detailed derivation, we find that the abelian gauge field and field strength are then related to $\vec{X}$ by
\begin{equation} \label{eq_abeliangauge}
A_z' = A_x' + i A_y' = \frac{\alpha (\vec{X} \cdot \vec{\mathcal{A}})}{x+X_3} , \quad B' = \vec{\nabla} \times \vec{A}' .
\end{equation}
Therefore, it is sufficient to solve for a periodic hidden constant of motion $\vec{X}$ satisfying 
\begin{equation} \label{eq_equationconstantmotion}
\left[ \vec{X} \cdot \vec{\mu} , \mathcal{D} \right]  = 0 .
\end{equation}

We now explicitly construct this constant of motion using the zero-energy states of the symmetry-protected Dirac cones at the $\kappa$ and $\kappa^\prime$ corners of TBG's Brillouin zone. 
More precisely, regardless of the value of $\alpha$, 
we are guaranteed two zero modes from the $C_2 T$ symmetry of the Hamiltonian~\cite{tarnopolsky2019origin}, which are pinned at the Brillouin zone corners in the presence of $C_3$ symmetry. 
Because $\mathcal{D}^\dagger({r}) = \mathcal{D}^*(-{r})$, the zero modes of $\mathcal{D}^\dagger$ are related to those of $\mathcal{D}$ by $C_2 T$ transformation: 
\begin{equation}
    \mathcal{D} |\psi^{}_{\kappa/ \kappa^{\prime}} ({r})\rangle = 0, \quad \mathcal{D}^\dagger |\psi^*_{\kappa/ \kappa^{\prime}} (-{r})\rangle = 0
\end{equation}
with $|\psi^{}_{\kappa, \kappa^{\prime}} (\mathbf{r})\rangle$ a layer-spinor that is fully polarized on the $A$ sublattice. 
From these definitions, it is clear that 
\begin{equation} \label{eq_definitionconstantofmotion}
    \vec{X}\cdot\vec{\mu} = |\psi^{}_{\kappa} ({r})\rangle \langle \psi^{*}_{\kappa} (-{r})| - |\psi^{}_{\kappa^\prime} ({r})\rangle \langle \psi^{*}_{\kappa^\prime} (-{r})|
\end{equation}
satisfy the commutation relation Eq.~\ref{eq_equationconstantmotion}. In App.~\ref{app_detailsinvertibletransfo}, we prove that $\vec{X}\cdot\vec{\mu}$ is traceless {provided that the relative phase between $\psi_\kappa$ and $\psi_{\kappa^\prime}$ is properly fixed}. We also find that the norm $x$ is uniform in space and equals the Dirac velocity $v_F(\alpha)$ around the $\kappa/\kappa'$ points for all $\alpha$. 

Away from the magic angle where $v_F(\alpha) \neq 0$~\cite{bistritzer2011moire}, the invertible transformation in Eq.~\ref{eq_transformeddirac} decouples the chiral TBG Hamiltonian into two independent Dirac Hamiltonian in an effective periodic magnetic field whose total flux through each unit cell needs to vanish to uphold the original translation symmetry of the model. 
The magic angle defined by $v_F(\alpha) = 0$ results in a rank-deficient constant of motion and a singular transformation $U$. 
As we will see below, this singularity physically corresponds to a divergence of the effective abelian magnetic field at a point where $(-2\pi)$ flux quantum is threaded. 
Well-behaved zero modes of the problem must vanish at this singular point, thereby screening this negative flux quantum. Since the total flux through the moiré unit cell vanishes, these zero modes effectively feel an average of $2\pi$ flux per moiré unit cell, leading to a flat band protected by the AS index theorem. 
We detail these behaviors below.

\subsection{Effective magnetic field} \label{ssec_effectivemagneticfield}

\begin{figure}
\centering
\includegraphics[width=\columnwidth]{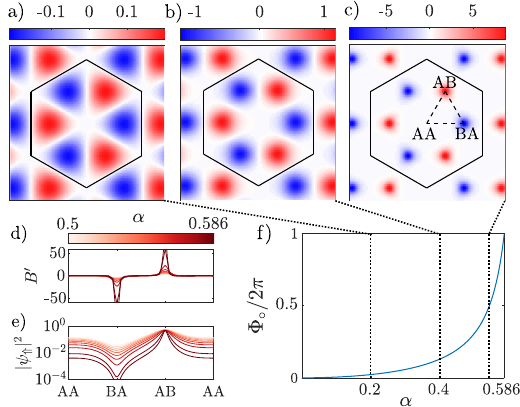}
\caption{
Effective magnetic field $B^\prime$ in real space at (a) $\alpha = 0.2$, (b) $\alpha = 0.4$ and (c) $\alpha = 0.55$. Symmetry guarantees that $B^\prime(-x, y) = - B^\prime(x, y)$. The $B'$ field becomes more and more localized as $\alpha$ gets closer to the first magic angle. (d) Wavefunction density $|\psi_{\Uparrow}|^2$ and (e) effective magnetic field $B^\prime$ along the high-symmetry point contour represented by a dashed line in (c) as a function of $\alpha$ (different colors). 
(f) Dimensionless flux of $B^\prime$ integrated over a small circle of radius $0.5 
/ |\kappa|$ around AB stacking points as a function of $\alpha$ up to the first magic angle. 
At the first magic angle, (d-f) show that the magnetic field diverges at points where $\pm 2\pi$ flux are threaded, and that the wavefunction develops a zero at the BA stacking points where $B'$ is infinite and negative. 
As a result, the system effectively screens one flux quantum and feels an effective positive magnetic flux of $2\pi$ per unit cell.
}
\label{fig_small_alpha}
\end{figure}

To give a more intuitive understanding of these statements, let us describe the evolution of the system as $\alpha$ increases from zero to the first magic angle. 
Fig.~\ref{fig_small_alpha} shows the magnetic field $B'$ defined in Eq.~\ref{eq_abeliangauge} for $\alpha = 0.2, 0.4,$ and $0.55$. 
Contrary to the model Eq.~\ref{eq_chirltbg} that is periodic on a $\sqrt{3}\times\sqrt{3}$-extension of the moir\'e unit cell (represented with a solid line hexagon in Fig.~\ref{fig_small_alpha}), the magnetic field has the same periodicity as the moir\'e lattice. 
{This is guaranteed by lattice translational symmetry, which demands that $\mathcal{D}^\prime({r} + {a}_{1, 2}) = \mathcal{D}^\prime({r})$ (see App.~\ref{app_detailsinvertibletransfo}).}
In addition, $C_{2y}T$ symmetry requires that $B^\prime(-x, y) = -B^\prime(x, y)$ and vanish along the both $y$-axis and the lines related to it by $C_{2x}$ and $C_{3z}$ and translation symmetries. 
Another consequence of $C_{2y}T$ is that the total flux of $B'$ vanishes in each moir\'e unit cell. For small values of $\alpha$, the field is positive in right-oriented triangles and negative in left-oriented ones. Such a simple structure is, however, very different at larger values of $\alpha$, above the first magic angle, where at each triangle, the effective field fluctuates dramatically, and changes sign rapidly.

Below the first magic angle, as $\alpha$ grows toward it, we can observe that the maximum of $B'$ strongly increases and gets confined around the AB and BA stacking points (Fig.~\ref{fig_small_alpha}a-c). Precisely at the first magic angle, $B' \to \infty$ at the AB point and diverges with an opposite sign at the BA point. This limit can be seen in Fig.~\ref{fig_small_alpha}d. 
This has important consequences for the zero-energy modes of $\mathcal{D}'$, as we now show. Let us focus on one Abelian component, labelled by $\Uparrow$. The zero mode satisfies the differential equation $( - 2 i \partial_{\bar{z}} + A_z' )\psi_{\Uparrow} = 0$, whose solutions are formally given by
\begin{equation} \label{eq_genericzeromode}
\psi_{\Uparrow} ({r}) = f(z) e^{- \rho ({r})} , \quad \nabla^2 \rho = B' , 
\end{equation}
with $f(z)$ a function of $z=x+iy$, which we leave unspecified for the moment. Here $\rho$ is the Poisson potential for the effective magnetic field~\cite{estienne2023ideal}. 
When $B'$ diverges, $\rho$ diverges with the same sign, yielding a node to the wavefunction $\psi_{\Uparrow}({r})$ at the position where $B' \to -\infty$. This is exactly what we observe in Fig.~\ref{fig_small_alpha}d-e through numerically solving the spectrum of $\mathcal{D}'$ at momentum $\kappa$.

Integrating the flux in a small circle around the AB point, we see that each of these divergences threads precisely one unit of flux through the system (Fig.~\ref{fig_small_alpha}f). 
To isolate the effects of these $B'$ singularities at AB and BA, we split the magnetic field at magic angles as 
\begin{equation} \label{eq_singularfluctuationmagneticfield}
B^\prime({r}) = 2\pi \sum_{t }\left[\delta({r} - {r}_{\rm AB} - {t}) -\delta( {r} - {r}_{\rm BA} - {t})\right] + \tilde B({r}) , 
\end{equation}
where $t$ runs over moir\'e Bravais lattice vectors. 
The zeros of $\psi_\Uparrow$ ($\psi_\Downarrow$) at the BA (AB) point effectively screen the negative flux quantum threaded by the magnetic field. As a result, the zero modes at the magic angle feel a non-zero net magnetic flux of $2\pi$ in each moiré unit cell and form a Dirac Landau level protected by the AS theorem.

The same singular behavior of $B'$ occurs at the AB and BA stacking points for all higher magic angles, as shown in Fig.~\ref{fig_compare_pseudo_field}a-c. 
For higher magic angles, however, there are multiple singular points in $B'$ away from these high-symmetry points. 
This foreshadows the pathological behaviors of the higher magic angles, which will map onto effective magnetic field with much greater fluctuations than the smooth and well-behaved field obtained at the first magic angles (Fig.~\ref{fig_compare_pseudo_field}d-f). 
At its core, this difference between first and higher magic angle can be traced back to the fact that the $SU(2)$ effective magnetic field $\mathcal{B} = \nabla \times \vec{\mathcal{A}} + i \vec{\mathcal{A}} \times \vec{\mathcal{A}}$ is --- in the original basis --- almost abelian for small $\alpha \lesssim 1$ where the first magic angle lies (because $\vec{\mathcal{A}} \times \vec{\mathcal{A}} \propto \alpha^2$ is parametrically smaller than $\nabla \times \vec{\mathcal{A}} \propto \alpha$), but it is strongly non-abelian for larger $\alpha \gtrsim 1$ where the higher magic appear. 
As a result, the basis transformation that abelianizes the Hamiltonian need to strongly fluctuate for higher magic angle, giving rise to more non-uniform and less smooth effective fields.

\subsection{Landau level and magnetic field at magic angles} \label{ssec_landaulevelsandeffectivemagicangles}

To show the connection to Landau level physics in a more formal manner, we explicitly construct the Landau level basis states using arguments originally developed in Ref.~\cite{tarnopolsky2019origin}, to which we refer for additional details. The decomposition of the magnetic field into a singular and fluctuating part carries over to the Poisson field and yield
\begin{equation}
e^{-\rho} = e^{- \tilde \rho }  \prod_{t} \left|\frac{{r} - {r}_{\rm BA} -{t}}{{r} - {r}_{\rm AB} - {t}}\right| , \quad \nabla^2 \tilde \rho = \tilde B . 
\end{equation}
The zeros resulting from the singular part of the magnetic field allows to choose the meromorphic function 
\begin{equation} \label{eq_Blochperiodicfunctions}
f_k(z) = e^{2\pi k_1 z} \dfrac{\theta_1 (z - z_{\rm BA} - k_x -i k_y | \omega )}{\theta_1 (z - z_{\rm BA}|\omega )}, 
\end{equation}
in Eq.~\ref{eq_genericzeromode} to build a zero mode $\psi_{\Uparrow, k}$ of $\mathcal{D}'$ with a well-defined momentum $k$. 
Here, $\theta_1$ represents the Jacobi theta function, and we have fixed the moir\'e lattice constant to one, and defined $b_{1,2} = \kappa_{1,2} - \kappa_0$ with which we have decomposed $k = k_1 b_1 + k_2 b_2$.

One could guess that a similar construction for the second abelian component of $\mathcal{D}'$ would give another band $\psi_{\Downarrow,k}$ of zero modes. However, it is not the case. The rank deficiency of $\vec{X} \cdot \vec{\mu}$ at the magic angles ensures that $\ket{\psi_{A,k}} = U[\psi_{\Uparrow, k}, 0]^T$ and $\ket{\psi_{A,k}'} = U[0, \psi_{\Downarrow, k}]^T$ are linear dependent once transformed back into the original layer-basis of the Hamiltonian. Here, $\psi_{\Uparrow, k}$ denotes the zero-energy wavefunction of the top abelian component of the decoupled Dirac operator $\mathcal{D}'$ that carries quasi-momentum $k$.
In other words, we recover the desired flatband counting for chiral TBG, with one exact flat band per spin, valley, and sub-lattice.

\begin{figure}
\centering
\includegraphics[width=1\columnwidth]{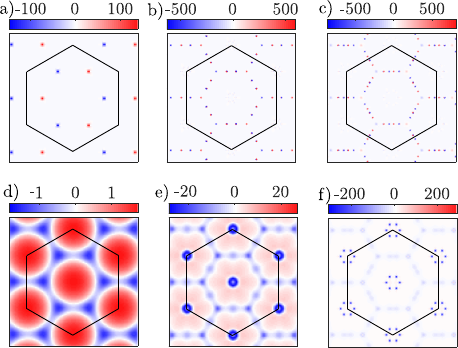}
\caption{The effective magnetic field $B^\prime$ at the (a) first, (b) second and (c) third magic angles. At all magic angles, $\delta$-function peaks are observed at the AB/BA stacking points, each carrying a flux of $2\pi$ in magnitude. Due to symmetry, the total flux within each unit cell remains zero. Panels (d)-(f) show the effective magnetic field $B^{\prime\prime}$ after applying the singular transformation (Eq.~\ref{eq:singular_trans}) for the first to third magic angles. The total flux in each unit cell now sums to $2\pi$, resulting in the formation of a flat band with an AS index.
}
\label{fig_compare_pseudo_field}
\end{figure}

To remove the singular component of the magnetic field and obtain an intuitive picture of the magnetic field effectively felt by $\ket{\psi_{A,k}}$, we include the singular part of the magnetic field leading to a node in the eigenstates' real-space density as follows. 
Instead of the almost everywhere invertible transformation $U$, we use a singular transformation $\Tilde{U} = U \Lambda(r) \mu_3$ to define
\begin{align}\label{eq:singular_trans}
\mathcal{D}^{\prime\prime} & = \Tilde{U}^{-1} \mathcal{D} \Tilde{U} = - 2 i \mu_0\partial_{\bar{z}} + A_z^{\prime\prime} \mu_3  , \\
A_z^{\prime\prime} & = A_z^\prime + 2i \partial_{\bar{z}}\ln \Lambda , \quad B^{\prime \prime} = \nabla \times A^{\prime\prime} , \notag
\end{align}
where choosing 
\begin{equation}
\Lambda ({r}) = \frac{\theta_1 (z-z_{\rm BA}|  \omega )}{|| U [1,0]^T ||}, 
\end{equation}
eliminates all singularities in the density of the eigenvectors.
We compare the original $B^{\prime}$ field with the non-singular and well-behaved $B^{\prime\prime}$ resulting from the singular transformation in  Fig.~\ref{fig_compare_pseudo_field}, which were first discussed in Ref.~\cite{ledwith2020fractional} for the first magic angle. 
Note that this singular transformation can only be carried out at the magic angle, where the $B^\prime$ field is itself singular, such that the zero-mode wavefunction have nodes at the BA stacking points.

The additional divergences in $B'$ of higher magic angle manifest in $B''$ as oscillations of larger and larger amplitudes (note the scales in Fig.~\ref{fig_compare_pseudo_field}d-f). 
Though all magic angles admit a Landau level description, we now argue and numerically demonstrate that these strong fluctuations prevent the protection from the index theorem to play any role in realistic settings.

\section{Effect of disorder} \label{sec_disorder}

In the previous section, we identified a hidden constant of motion decoupling the flat bands in chiral TBG, endowing them with an analytical index. 
Here, we phenomenologically argue how the flat bands of chiral TBG respond to various forms of disorder. 
Our analysis unveils that only the first magic angle successfully harnesses the topological protection offered by the index theorem. 
To see this, we classify the disorder channels from the most to the least detrimental for the flat bands in Sec.~\ref{ssec_hierarchydisorder}. 
Our results, summarized in Tab.~\ref{tab_summaryhierarchy}, are then substantiated by numerical simulations both in (Sec.~\ref{sec_numerics}) and away from (Sec.~\ref{ssec_disordernonchirallimit}) the chiral limit.

\begin{table*}
\centering
\begin{tabular}{l||l|l||l}
disorder channel & first magic angle & higher magic angles & example \\ \hline \hline 
chiral $\sigma_{1/2} \mu_0$ & protected &  protected & homo-strain \\ \hline 
chiral $P(\vec{X})$ & \begin{tabular}[c]{@{}l@{}} suppressed by $P(X)$ acting locally \end{tabular} & not protected & hetero-strain \\ \hline 
resonant non-chiral & \begin{tabular}[c]{@{}l@{}} suppressed by wavefunction localization in $k$ \end{tabular} & not protected  & $AB$-tunneling \\ \hline 
off-resonant & suppressed by spectral gap & \begin{tabular}[c]{@{}l@{}} not protected, gap squeezing \end{tabular}  & $AA/BB$-tunneling  
\end{tabular}
\caption{Effect of the different disorder channels on the flat bands. 
The first two lines describe disorder that anticommute with the chiral symmetry in the decoupled abelian basis of the Hamiltonian, to which the flat bands are perfectly immune thanks to the AS theorem. 
We split the chiral disorder channels into those independent of the magic angle -- the $\sigma_{1/2} \mu_0$ homo-strain of the first line -- and the fine-tuned channel $P(\vec{X})$ that explicitly depend on the constant of motion $\vec{X}$ at each magic angle. 
This fine-tuned disorder only has a large overlap with physical disorder realizations that carry no momentum on average at the first magic angle where $P(\vec{X})$ acts locally (see Fig.~\ref{fig_normXmomentumspace}). 
The third line describes resonant disorder that directly couples the two sublattice polarized flat bands, and encompasses variations of the $AB$-tunneling amplitude. 
Only the first magic angle shows strong protection against this type of disorder (see also the numerical results of Sec.~\ref{sec_numerics}) due to the strong localization of $\vec{X}$ in momentum space that mitigates the effects of all the disorder's high-harmonics (Fig.~\ref{fig_normXmomentumspace}).
Finally, the last line concerns off-resonant terms connecting the flat bands to degrees of freedom above the spectral gap. 
The strong fluctuations of the effective magnetic field at higher magic angle (Fig.~\ref{fig_compare_pseudo_field}) yield an exponentially small gap for the higher magic angles, as first proved in Ref.~\cite{becker2023magic} based on generic argument for non-uniform Landau problems.
}
\label{tab_summaryhierarchy}
\end{table*}

\subsection{Hierarchy between disorder channels} \label{ssec_hierarchydisorder}

We split the various types of disorder susceptible to broadening the flat bands of chiral TBG into three categories: the chiral channels anticommuting with the chiral operators of the two decoupled abelian components of the Hamiltonian, the resonant scattering processes directly coupling the two sublattice polarized flat-bands, and the off-resonant terms scattering particles from the flat-band to degrees of freedom above or below the spectral gap. 
We here provide a physical discussion of the effects of these different disorder channels on the flat bands. 

\subsubsection{Chiral disorder}  

The Atiyah-Singer index ensures the stability of the flat bands against any perturbations anticommuting with the chiral symmetry $\sigma_3$ in each of the decoupled abelian components of the Hamiltonian separately. 
In other words, after the basis transformation of Eq.~\ref{eq_transformeddirac}, the system can sustain any kind of disorder in the $\sigma_{1/2} \mu_0$ and $\sigma_{1/2} \mu_3$ channels. 
The first of these operators is independent of the layer-basis and takes the same form in the original degree of freedom of the problem. 
In the physical basis of Eq.~\ref{eq_chirltbg}, disorder in the $\sigma_{1/2} \mu_0$ channel describes homo-strain, against which all the magic angle are protected by virtue of the AS index theorem -- as summarized in the first line of Tab.~\ref{tab_summaryhierarchy}.

\begin{figure}
\centering
\includegraphics[width=0.9\columnwidth]{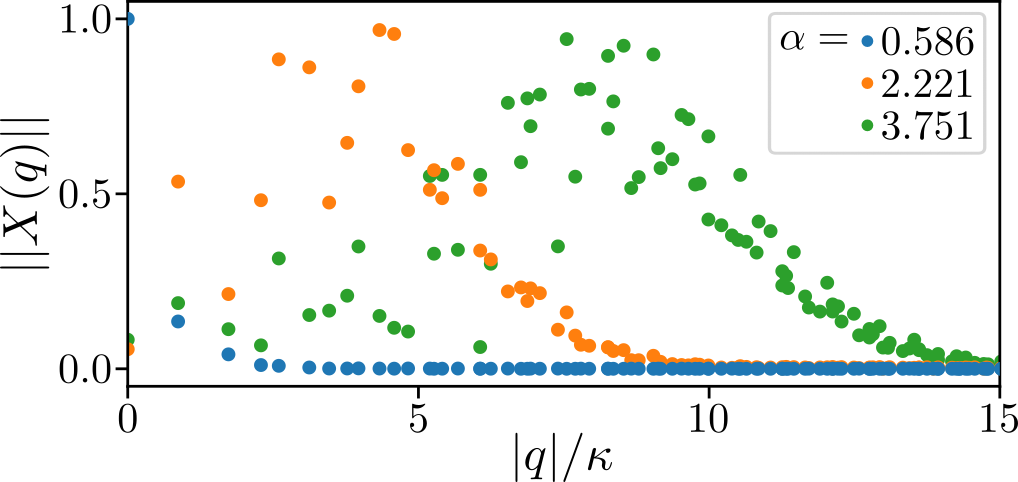}
\caption{Norm $||X(q)|| = [\sum_i |X_i(q)|^2]^{1/2}$ in momentum space for the first (blue), second (orange), and third (orange) magic angles. Only at the first magic angle is $P(\vec{X})$ dominated by its $q=0$ component and therefore acts locally. For visibility, we only included momenta with $q_x, q_y \geq 0$.
}
\label{fig_normXmomentumspace}
\end{figure}

In contrast, transforming $\sigma_{1/2} \mu_3$ back to the original basis gives 
\begin{equation}
P(\vec{X}) = \begin{bmatrix} 0 & \vec{X}\cdot\vec{\mu} \\ \vec{X}^*\cdot\vec{\mu} & 0 \end{bmatrix} ,
\end{equation}
a fine-tuned disorder channel that explicitly depends on $\vec{X}$, and hence on the details of the abelianization at each magic angle. 
While this disorder channel contains fine-tuned combinations of layer Pauli operators, it can still efficiently protect the band flatness provided realistic forms of disorder largely overlap with it. 
To determine whether typical disorder realization can, in principle, overlap with $P(\vec{X})$, it is instructive to look more closely at the structure of the constant of motion $\vec{X}$ at the various magic angles. 
In Fig.~\ref{fig_normXmomentumspace}, we display $||X(q)|| = [\sum_i |X_i(q)|^2]^{1/2}$ the norm of the Fourier transform of $\vec{X}$. 
We observe that it is sharply peaked at $q=0$ for the first magic angle, highlighting that $P(\vec{X})$ acts locally, whereas $||X(q)||$ exhibits maxima at finite momenta for higher magic angles. 
As a result, random disorder realization carrying no zero average momentum will only carry high overlap with $P(\vec{X})$ at the first magic angle. 
This suppression of the disorder strength due to finite overlap with $P(\vec{X})$ concerns all channels contained in it and includes both hetero-strain and variations of the $AB$-tunneling amplitude.

\subsubsection{Non-chiral resonant disorder}  

Let us now consider disorder channels that do not necessarily anti-commute with the chiral symmetries in the decoupled abelian basis, but nevertheless yield direct coupling between the two sublattice flat bands. 
These disorder channels are carried by the $\sigma_{1/2}$ orbital Pauli matrices and hence anticommute with the original chiral operator $\sigma_3$ of the original chiral Hamiltonian of TBG. 
For the sake of concreteness, we focus on the $\sigma_{1/2} \mu_{1/2}$ channels, which physically correspond to perturbations to the $AB$ tunneling amplitude that modify the effective gauge potential $\mathcal{A}_z$ by introducing higher moir\'e harmonics and can be due, for instance, to out-of-plane lattice relaxation.

These resonant disorder channels induce a first-order broadening of the band proportional to a matrix element connecting the two sublattice polarized flat bands by some high moir\'e harmonics. 
This matrix element is small (resp. large) if the dominant disorder harmonics is large (resp. small) compared to the extent of the flat band wavefunctions in momentum space, such that the disorder matrix element involves momentum points with weak (resp. large) amplitude of wavefunctions (App.~\ref{subapp_momlock}). 
Indeed, the envelope of the flat band wavefunctions $\ket{\psi_{A,k}}$ in momentum space is determined by the transformation $U$, which at the magic angle is proportional to $\vec{X} \cdot \vec{\mu}$ (see Sec.~\ref{ssec_landaulevelsandeffectivemagicangles}). The functional form of the $\vec{X}$ vector therefore strongly influences the effects that disorder has on the flat bands.
As a result, the flatness of the band is protected against most forms of tunneling disorder provided $\vec{X}$ is strongly localized in momentum space, and strongly susceptible to broadening when $\vec{X}$ has support over a large number of harmonics in momentum space. 
In Fig.~\ref{fig_normXmomentumspace}, we show that $\vec{X}$ is strongly localized in momentum space only at the first magic angle, which again highlights the importance of the locality of the constant of motion $\vec{X}$ in the mitigation of disorder and protection of the band flatness. This is summarized in the second line of Tab.~\ref{tab_summaryhierarchy}.

More formally, the magic angles can be obtained from the eigenvalues of the operator $\Lambda =  \hat{D}^{-1} \hat{T}_+ \hat{D}^{-1} \hat{T}_-$ (see App.~\ref{subapp_momlock})~\cite{becker2022mathematics}, where $T_\pm = \mathcal{A}_z(\pm r)$ and $\hat{D}^{-1}$ is the Dirac propagator. Disorder modifies the tunneling terms as $T \rightarrow T + \delta T$. We can now use perturbation theory in $\delta T$ to understand the robustness of the magic angle.
The perturbation $\delta T$ transfers momenta from $k$ to $k+q$, and the propagator penalizes this transfer by a factor $1/(k+q)$.
As a result, for a tunneling disorder with correlation length $\xi$, wavefunctions that are confined in momentum-space, with size $k_\xi \sim 2\pi/\xi$, are less prone to broadening. 
As shown in the App.~\ref{app_detailsnumerics} but also explicit in Fig.~\ref{fig_compare_pseudo_field}d-f, the wavefunction of the first magic angle is the most confined in momentum space and is, hence, the most robust. 
In our numerical simulation presented below, we find that the position of the first magic angle remains remarkably robust against these perturbations, while the position of higher magic angles is extremely sensitive to their effects.

\subsubsection{Non-chiral non-resonant disorder} 

Disorder channels of this kind connect the flat bands to degrees of freedom away from zero energy. Included in this category is, for instance, inter-layer tunneling that preserves the sublattice. If the disorder strength is small compared to the spectral gap between the flat bands and the rest of the spectrum, the disorder only affects the flat band’s bandwidth in second order. This provides an effective protection of the flat band against off-resonant scattering.

In Ref.~\cite{becker2023magic}, it was shown that the spectral gap between the flat bands and the rest of the spectrum decays exponentially as a function of $\alpha$ as the latter increases. 
Using the decoupling introduced in Sec.~\ref{sec_theory}, we interpret this as a result of the exponential squeezing of the spectrum of Landau-like operators as a function of the amplitude of the magnetic field's fluctuations~\cite{hormander1960differential,duistermaat1973global,zworski2001remark,becker2024absence}. 
Fig.~\ref{fig_compare_pseudo_field}d-f illustrate that these fluctuations grow rapidly with $\alpha$: they increase more than tenfold with each successive magic angle, and are only smaller than the average magnetic field at the first magic angle. 
This rapid increase leads to a sharp collapse of the spectral gap above the flat bands: at the first three magic angles $\alpha = 0.586, 2.221, 3.751$ the spectral gap respectively read $0.29$, $0.059$ and $0.0027$ in units of the typical moir\'e kinetic energy $|v\kappa|$ (that itself decreases linearly with $\theta$).  
The increasing relevance of the non-uniform part of the magnetic field, leading to an exponential reduction of the gap as a function $\alpha$ for large $\alpha$, was recently used to prove the impossibility of observing higher magic angles in TBG~\cite{becker2024absence}. 
This is summarized in the last line of Tab.~\ref{tab_summaryhierarchy}.

To summarize (see Tab.~\ref{tab_summaryhierarchy}), the first magic angle possesses intrinsic robustness to a broad class of disorder channels thanks to its topological origin, large gap, and the localization of its wavefunctions that originates from its well-behaved constant of motion $\vec{X}$. On the other hand, higher magic angles are expected to strongly broaden in the presence of any disorder except homo-strain. We now probe this qualitative difference numerically.

\subsection{Numerical simulations} \label{sec_numerics}

To bolster these arguments, we numerically simulate the effect of disorder in the chiral model of TBG. Concretely, we introduce a $C_3$ symmetric disorder by adding higher-order harmonics to the inter-layer tunneling potential. This amounts to the substitution $\mathcal{A}_z \to \mathcal{A}_z + \delta \mathcal{A}_z$ in Eq.~\ref{eq_chirltbg} with
\begin{align} \label{eq_highharm}
& \delta \mathcal{A}_z = \sum_{m = 1}^{m_c} \mu_m ( a_{m,z}^{(1)} \mu_1 + a_{m,z}^{(2)} \mu_2), \\
& a_{m,z}^{(1)} + i a_{m,z}^{(2)} = \sum_{n} e^{i (\kappa_{m,n} \cdot r +\phi_{m, n})} , \notag
\end{align}
where $\kappa_{m,0}$ is the $m$-th smallest vector in the first trine of the plane that connects the two corners $\kappa$ and $\kappa'$ of the moir\'e Brillouin zone (up to reciprocal lattice vectors), $\kappa_{m,n}$ denotes its rotation by $2n\pi/3$, and $\mu_m$ stands for the amplitude of the $m$-th harmonic. The $\phi_{m, n}$ phases are included such that $\delta \mathcal{A}_z$ does not break $C_3$ and the $C_{2y}$ symmetry of Eq.~\ref{eq_chirltbg}. The precise definition of all these terms is given in App.~\ref{app_detailsnumerics}. We only consider the first $m_c = 10$ harmonics and discretize $\mathcal{D}$ on a momentum-space lattice containing all plane waves with wavevectors of norm smaller than a certain radius $R_k$.

\begin{figure}
\centering
\includegraphics[width=\columnwidth]{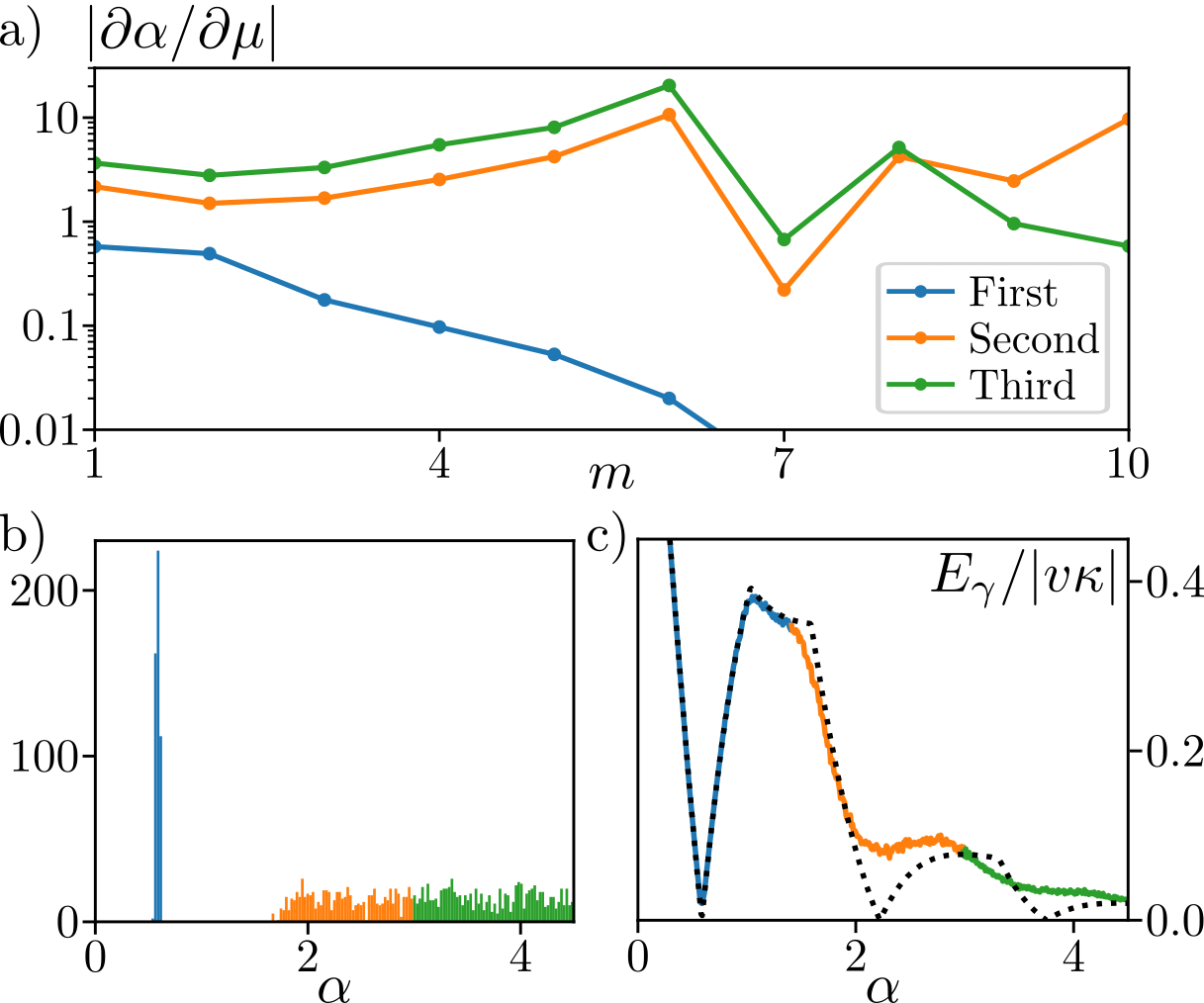}
\caption{a) Variation $|\partial \alpha / \partial \mu|$ of the first (blue), second (orange), and third (green) magic angle computed for a small $\delta \mu = 0.02$ perturbation in the $m$-th harmonics of the $AB$ inter-layer tunneling (see Eq.~\ref{eq_highharm}). To obtain converged results for large $m$, we used a large truncation radius $R_k = 25 \sqrt{3} |\kappa|$. 
b) Histogram of the magic angles for $500$ realization of the inter-layer tunneling disorder with all $\mu_m$ drawn uniformly and normalized as $\sqrt{\sum_m \mu_m^2} = \delta\mu = 0.1$. 
c) $E_\gamma$, the energy of the top active band crossing zero energy in chiral TBG at the center of the moir\'e Brillouin zone ($\gamma$), averaged over $500$ disorder realizations for the same distribution of $\mu_m$ as in (b). The dashed line shows the result in the absence of disorder. 
$R_k = 25\sqrt{3}|\kappa|$ for (b) and $R_k = 10\sqrt{3}|\kappa|$ for (c) is used. Colors are provided as a guide to the eye, allowing us to identify the regions closest to 0.586 (blue), 2.221 (orange), and 3.751 (green), which correspond to the magic angles in the absence of disorder.}
\label{fig_sus_exp}
\end{figure}

For the considered $C_3$-preserving inter-layer tunneling disorder, the chiral model of TBG still possesses exact magic angles where the bands at zero energy become perfectly degenerate~\cite{sheffer2023symmetries,becker2022mathematics}. 
The precise value of these magic $\alpha$ can be efficiently determined using the method of Refs.~\cite{becker2022mathematics,becker2023magic}, where all magic angles are determined as eigenvalues of the operator $\Lambda$ (see App.~\ref{app_detailsnumerics} for details).
We first use this method to characterize the sensitivity of the magic angles to the different harmonics in Eq.~\ref{eq_highharm}. 
For this purpose, we choose one specific harmonic $m$ and compute the change $\delta \alpha$ in the first, second, and third magic angles when $\mu_m = \delta \mu$ is set to a small but finite value (keeping all other $\mu_{n\neq m}=0$). 
For small disorder strength, this corresponds to taking the derivatives $\partial \alpha / \partial \mu$ of the eigenvalues of $\Lambda$  with respect to $\mu_m$. 
We numerically evaluate this derivative using a small enough $\delta\mu=0.02$, and plot our results for $|\partial \alpha / \partial \mu|$ in Fig.~\ref{fig_sus_exp}a. 
We observe that the $\alpha$ deviations in the first magic angle exponentially decay with $m$ and are negligible for $m > 5$, indicating that the first magic angle is robust against microscopic details on scales smaller than the moir\'e length. 
In contrast, the deviations for the higher magic angle can reach up to ten times the amplitude of the added perturbation and do not decay with $m$. 
This suggests that the higher magic angles are easily affected by microscopic disorder at all scales down to graphene's lattice constant. 
Perturbations on those scales, such as lattice relaxations, are inevitable and necessarily spoil the prediction for higher magic angles~\cite{tarnopolsky2019origin,becker2022mathematics}. 
Perturbations on scales larger than the moir\'e length are also studied in App.~\ref{app_detailsnumerics}.

To understand the physical implications of the difference in $\delta \alpha$ for first and higher magic angles observed in Fig.~\ref{fig_sus_exp}a, we consider realistic disorder realization containing all harmonics $m\leq m_c$ with amplitudes $\mu_m$ drawn uniformly in $[-1, 1]$ with the overall normalization $\sqrt{\sum_m \mu_m^2} = \delta\mu = 0.1$.  
We plot the distribution of the first three magic angles for 500 realizations of this disorder in Fig.~\ref{fig_sus_exp}b, which shows a similar trend as Fig.~\ref{fig_sus_exp}a: the distribution of magic $\alpha$ is sharply peaked around the original value 0.586 for the first magic angle. At the same time, it is extremely broad and featureless for higher magic angles. 
To interpret this result, picture a sample in which different patches of the sample (different unit cells) are characterized by slightly different atomic arrangements and lattice relaxation profiles that locally change the values of the $\mu_m$. 
We can assign an effective magic angle $\alpha$ to each of these regions. 
When TBG is close to the first unperturbed magic angle, the distribution of these magic $\alpha$ is peaked, and the magic condition is almost simultaneously satisfied for all patches (see Fig.~\ref{fig_sus_exp}b). 
This is not true for higher magic angles, such that the density of states at zero energy is smeared in many patches, yielding a strong broadening of the original flat bands.

We can estimate the flat band broadening by computing $E_\gamma$ the average energy of the topmost band at the center of the moir\'e Brillouin zone ($\gamma$ point) for 500 realizations of the realistic disorder pattern of Eq.~\ref{eq_highharm} for a fixed twist angle, \textit{i.e.} as a function $\alpha$. 
The results presented in Fig.~\ref{fig_sus_exp}c pictorially capture our previous statement: the higher magic angles are completely washed out due to their high sensitivity to disorder. 
The average $\gamma$-point energy only displays an apparent non-monotonic behavior with an almost vanishing bandwidth at the first magic angle, which we have seen to be protected against disorder (Tab.~\ref{tab_summaryhierarchy}).

Our numerical calculations clearly show that only the flat bands at the first magic angle retain a large density of states in the presence of disorder. 
As explained in Tab.~\ref{tab_summaryhierarchy}, this is due to the combination of the topological protection from the AS theorem {to local chiral symmetric disorder}, large spectral gap, and strong localization of the wavefunctions in the flat band. 
Higher magic angles are sensitive to lattice scale details and should be interpreted as accidental artifacts of the continuum model rather than experimentally observable features.

\subsection{Away from the chiral limit} \label{ssec_disordernonchirallimit}

\begin{figure}
\centering
\includegraphics[width=\columnwidth]{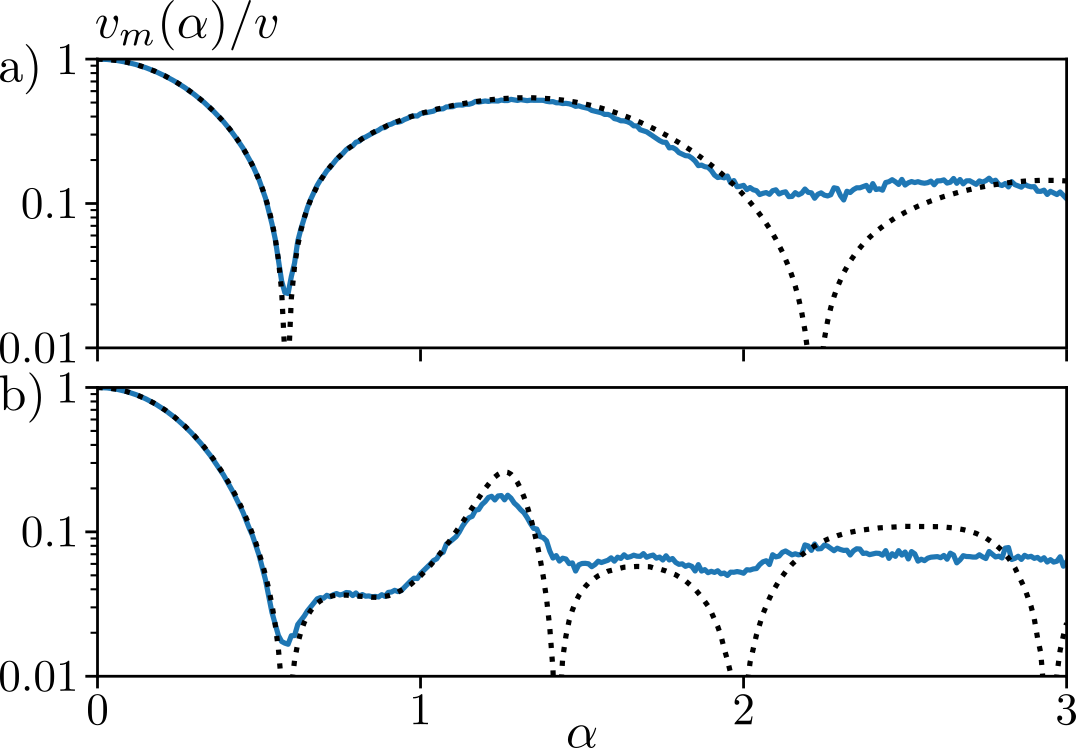}
\caption{Fermi velocity at the corner of the moir\'e Brillouin zone $v_m (\alpha)$ averaged over $500$ disorder realizations for the same distribution of $\mu_m$ as in Fig.~\ref{fig_sus_exp}, both in the chiral limit (a) and the realistic value for $AA$ tunneling strength $w_0= 0.8w$ (b). 
To obtain similar results at the first magic angle, allowing for a fair comparison, we have used $\delta\mu =0.1$ for the chiral limit (a) and $\delta\mu=0.2$ in the realistic regime (b). 
The dashed lines show the same quantity in the absence of disorder. We have normalized the results by the Fermi velocity of graphene $v$ and shown them on a log scale to better discern the effect of the weak disorder.}
\label{fig_awaychiral}
\end{figure}

We finally check whether our results, which find their roots in the chiral limit of TBG, also hold in the more realistic situation where non-zero $AA$ tunnelings are included in the Bistritzer-MacDonald model~\cite{bistritzer2011moire}. 
When the $AA$ tunneling amplitude $w_0$ is non-zero, we cannot identify the magic angle with the point where the $\gamma$-point energy vanishes, as we did in Fig.~\ref{fig_sus_exp}c, since the model never features perfectly flat bands. 
Instead, we locate the magic angle as the points for which the Fermi velocity at the corner of the moir\'e Brillouin zone vanishes~\cite{sheffer2023symmetries,bistritzer2011moire}. 
We have numerically computed this Fermi velocity $v_m (\alpha)$ averaged over 500 realizations of the generic disorder pattern described above as a function of $\alpha$, both in the chiral limit $w_0=0$ and for the realistic value $w_0= 0.8 w$~\cite{fang2019angle}. 
Our results are presented in Fig.~\ref{fig_awaychiral}, where, for better comparison, we have chosen $\delta\mu =0.1$ for the chiral limit and $\delta\mu=0.2$ in the realistic regime, leading to the similar average velocity at the first magic angle. 
In both the chiral and realistic (nonchiral) limits, disorder barely affects the Fermi velocity  (blue lines) from its clean values (dotted lines) as long as we are in the abelian regime ($\alpha \lesssim 1$). In the non-abelian regime, the average velocity deviations become significant.
Consequently, only the first magic angle still strongly suppresses the Fermi velocity $v_m (\alpha)$. In contrast, all higher magic angles are washed out by disorder -- we interpret this as a direct consequence of the robustness granted by the AS topological index described in the second line of Tab.~\ref{tab_summaryhierarchy}.

To understand why a stronger disorder strength is necessary for the realistic limit ($\delta \mu = 0.2$) to obtain the same average value of $v_m(\alpha)$ at the first magic angle than in the chiral limit ($\delta \mu =0.1$), we should take a closer look at the effect of the non-chiral part of the disorder. More precisely, it is known that the wavefunction of the flat bands is more localized near the $AA$ region in the realistic limit $w_0 = 0.8 w$ than in the chiral limit~\cite{tarnopolsky2019origin}. Therefore, the third line of Tab.~\ref{tab_summaryhierarchy} qualitatively captures the better protection of the flat band against inter-layer tunneling disorder.

\section{Implications for TMDs} \label{sec_tmd}

\begin{figure}
\centering
\includegraphics[width=\columnwidth]{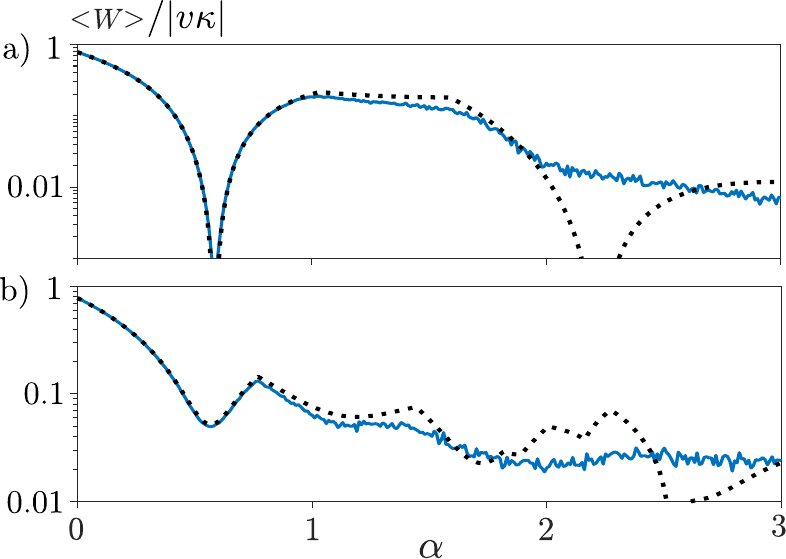}
\caption{Bandwidth of the lowest band of twisted TMDs modeled by a mass term $m = 0.25$ for a) $w_0 = 0$ and b) $w_0 = 0.465 w_1$
averaged over $50$ disorder realizations  of strength $\delta \mu = 0.1$ for the same distribution of $\mu_m$ as in Fig.~\ref{fig_sus_exp}, as a function of $\alpha$. The dashed line shows the same quantity in the absence of disorder. For b) we use the realistic parameter for WSe$_2$ \cite{crepel2023chiral}.
}
\label{fig_TMD}
\end{figure}

While TBG has been the main focus up to now, let us expand on the implications of our results to twisted TMDs, which share the same chiral limit as TBG (Eq.~\ref{eq_chirltbg}) up to the addition of a mass term that describes the gap of the semiconducting monolayers~\cite{crepel2023chiral}. 
Because they are sublattice polarized and, therefore, eigenstates of the mass term, the flat bands of the chiral model at the magic angle remain completely flat under the introduction of the mass term but shift to non-zero positive/negative energy to describe the topmost valence and bottom-most conduction moir\'e bands of the system. 
Most of our analysis, particularly the decoupling into abelian components from Sec.~\ref{sec_theory}, goes through in the presence of this mass term. 
To show this, we calculate the averaged bandwidth of the lowest band of twisted TMDs with a mass term as a function of $\alpha$ in Fig.~\ref{fig_TMD}. One clearly sees that while the flatness of the first magic angle is robust, that of higher magic angles is destroyed by disorder.
Away from the chiral limit, the resulting flat band will acquire some dispersion due to additional orbital-preserving terms in the Hamiltonian but remain robust against disorder if the results of Fig.~\ref{fig_awaychiral} can be generalized to twisted $\rm MoTe_2$ homobilayers. 

Our main result is that magic angles exists in twisted TMDs when their low-energy degrees of freedom are described by gapped Dirac cone with concentrated Berry curvature at the $K/K'$ points, and that the corresponding flat bands carry a topological protection against disorder as long as the effective magnetic field obtained is well-behaved, which occurs in the abelian limit $\alpha \lesssim 1$ where the moir\'e tunneling in the original Hamiltonian lead to an almost abelian gauge field from the start (see discussion at the end of Sec.~\ref{ssec_effectivemagneticfield}). 
Our numerical results presented in Figs.~\ref{fig_sus_exp}~-~\ref{fig_TMD} suggest that it is indeed the case.
Taking $\alpha = 1$ as the limit for topological protection, and using the effective Fermi velocity of graphene and $\rm MoTe_2$~\cite{PhysRevB.88.085433}, we expect the protection across all angles $\theta > 0.7^\circ$ in TBG and $\theta > 2^\circ$ for twisted $\rm MoTe_2$ homobilayers. In particular, the first magic angle of twisted $\rm MoTe_2$ at $\theta \sim 3.2^\circ$ is topologically protected~\cite{crepel2023chiral}.

The mass term and the strong spin-orbit coupling of TMDs~\cite{xiao2012coupled} strongly suppress all resonant non-chiral disorder -- the only ones that spoil the protection of the flat bands, see Tab~\ref{tab_summaryhierarchy} -- as they fully lift the degeneracy of the flat band to leave a single topologically protected valence flat band per spin flavor. 
Twisted TMDs, therefore, are ideal hosts for delicate correlated topological phases of matter. They host weakly dispersing Chern bands fully protected against all forms of disorder thanks to the combinations of the AS topological index theorem and their spectral gap.

We also note that other approaches aim at interpreting the physics of twisted TMDs as particles moving in an abelian gauge field~\cite{wu2019topological,morales2023magic,crepel2024bridging, li2024contrasting,Kang2023,Vafek2023}. 
Our work precisely determines in which regimes and for which angles such a picture is valid.

\section{Conclusion}

In this work, we have demonstrated that the flat bands appearing at the magic angles in twisted graphene and transition metal dichalcogenides bilayers, both in the chiral limit, can be understood as originating from Dirac particles evolving in an abelian gauge field. 
Thanks to a topological index theorem, these zero-energy Dirac Landau levels become perfectly flat and topologically protected against disorder when the system's moir\'e length and effective magnetic length match. 
Due to the pathological nature of the magnetic field at all but the first magic angle, we find that the topological protection offered by the index theorem can only be harnessed when the inter-layer tunneling is small enough compared to the typical Dirac kinetic energy at the moiré scale. 
We estimate the topologically protected region to span over all twist angles $\theta > 0.7^\circ$ for twisted bilayer graphene and $\theta > 2^\circ$ for twisted $\rm MoTe_2$. 
Only the first magic angle of both systems lies in this range, offering novel analytical insights into why delicate correlated phases such as fraction Chern insulators can be stabilized in these disordered systems.

\section{Acknowledgement}  Research on topological properties of moir\'e materials is supported as part of Programmable Quantum Materials, an Energy Frontier Research Center funded by the U.S. Department of Energy (DOE), Office of Science, Basic Energy Sciences (BES), under award DE-SC0019443.  The Flatiron Institute is a division of the Simons Foundation. N.R. acknowledges support from the QuantERA II Programme which has received funding from the European Union’s Horizon 2020 research and innovation program under Grant Agreement No 101017733.

\bibliography{RefsTopoFB}

\onecolumngrid \appendix \newpage

\section{Abelianization} \label{app_abelianization}

We here assume that $\mathcal{D}$ admits a constant of motion of the form $\vec{X} \cdot \vec{\mu}$, \textit{i.e.} $[\vec{X} \cdot \vec{\mu}, \mathcal{D}] = 0$, and 
derive calculate the specific form of the transformed Dirac operators $\mathcal{D}' = U^{-1} \mathcal{D} U$ with
\begin{equation}\label{eq:Utransformation}
U = U^{-1} = \frac{1}{\sqrt{N}} (\vec{X}\cdot \vec{\mu} + x \mu_3) , \quad N = 2 x (x + X_3) .
\end{equation}
A direct calculation gives 
\begin{equation} \label{appeq_transfohamiltonian} \begin{split}
\mathcal{D}' & = U^{-1} \mathcal{D} U , \\
& = \frac{1}{N} [(\vec{X}\cdot \vec{\mu}) \mathcal{D}(\vec{X}\cdot \vec{\mu}) + x^2 \mu_3 \mathcal{D} \mu_3] + \frac{x}{N} [\mu_3  \mathcal{D} (\vec{X}\cdot \vec{\mu}) + (\vec{X}\cdot \vec{\mu}) \mathcal{D} \mu_3] + \sqrt{N} (-2i\partial_{\bar{z}} N^{-1/2}) \mu_0 \\ 
& = \frac{1}{N} [ \underbrace{(\vec{X}\cdot \vec{\mu})^2}_{x^2 \mu_0} \mathcal{D} + x^2 \mu_3 \mathcal{D} \mu_3] + \frac{x}{N} [\mu_3 (\vec{X}\cdot \vec{\mu}) \mathcal{D}  + (\vec{X}\cdot \vec{\mu}) \mathcal{D} \mu_3] + \frac{i}{N} (\partial_{\bar{z}} N) \mu_0 \\ 
& = \frac{x^2}{N} [2\cdot (-2i\partial_{\bar{z}}) \mu_0] + \frac{x}{N} [ \underbrace{\{\mu_3 , \vec{X}\cdot \vec{\mu}\}}_{2 X_3} (-2i\partial_{\bar{z}})  + \underbrace{[\mu_3 , \vec{X}\cdot \vec{\mu}]}_{2i(\hat{u}_3\times \vec{X}) \cdot \vec{\mu}} (\alpha \vec{\mathcal{A}} \cdot \vec{\mu})] + \frac{2 i x}{N} (\partial_{\bar{z}} X_3) \mu_0 \\
& = \frac{2 x (x+X_3)}{N} (-2i\partial_{\bar{z}}) \mu_0 -  \frac{2 i x}{N} [(\partial_{\bar{z}} X_3) \mu_0 + i \alpha (\vec{X} \cdot \vec{\mathcal{A}}) \mu_3] + \frac{2 i x}{N} (\partial_{\bar{z}} X_3) \mu_0 \\
& = (-2i\partial_{\bar{z}}) \mu_0 + \frac{\alpha (\vec{X} \cdot \vec{\mathcal{A}})}{x+X_3} \mu_3 ,
\end{split} \end{equation} 
where the first line is a simple re-ordering, the second line uses the fact that $(\vec{X}\cdot \mu)$ commutes with $\mathcal{D}$, the third uses that $\vec{\mathcal{A}}$ has no $\mu_3$ component such that $\mu_3 (\vec{\mathcal{A}} \cdot \vec{\mu}) \mu_3 = - (\vec{\mathcal{A}} \cdot \vec{\mu})$, the fourth and fifth simplify the products of Pauli matrices. This represents two decoupled copies subject, each subject to an abelian $U(1)$ gauge field 
\begin{equation}\label{eq:U1gaugepotential}
\pm A'_z= \pm \frac{\alpha (\vec{X} \cdot \vec{\mathcal{A}})}{x+X_3} ,
\end{equation}
that is opposite for the two copies.

\section{Constant of motion} \label{app_detailsinvertibletransfo}

In this appendix, we show that the constant of motion $\vec{X} \cdot \vec{\mu}$ defined in Eq.~\ref{eq_definitionconstantofmotion} is traceless, of uniform norm, and reproduces Eq.~\ref{eq_transformeddirac}.

In order to construct $\vec{X} \cdot \vec{\mu}$ as in Eq.~\ref{eq_definitionconstantofmotion}, we need the existence of the two zero modes $|\psi^{}_{\kappa} (\mathbf{r})\rangle$ and $|\psi^{}_{\kappa^\prime} (\mathbf{r})\rangle$, which is guaranteed by $C_2 T$ and $C_3$ symmtries \cite{tarnopolsky2019origin}. If $C_3$ symmetry is broken in the system, there will still be at least two zero modes which can be used to construct the constant of motion, albeit not fixed at $\kappa$, $\kappa^\prime$ anymore. We assume the presence of $C_3$ throughout our derivation for simplicity.

We choose the relative phase between $|\psi^{}_{\kappa} ({r})\rangle$ and $|\psi^{}_{\kappa^\prime} ({r})\rangle$ such that $\operatorname{Tr} \vec{X} \cdot \vec{\mu}$ vanishes. This is because
\begin{equation}
    |\operatorname{Tr} (|\psi^{}_{\kappa} ({r})\rangle \langle \psi^{*}_{\kappa} (-{r})|) | = |\operatorname{Tr} (|\psi^{}_{\kappa^\prime} ({r})\rangle \langle \psi^{*}_{\kappa^\prime} (-{r})|) | = v_F(\alpha)
\end{equation}
is an integral of motion which is independent of $\mathbf{r}$ \cite{tarnopolsky2019origin}. It only depends on $\alpha$ and corresponds to the Fermi velocity at the two Moire Dirac cones. Therefore, $\operatorname{Tr} (|\psi^{}_{\kappa} ({r})\rangle \langle \psi^{*}_{\kappa} (-{r})|)$ and $\operatorname{Tr} (|\psi^{}_{\kappa^\prime} ({r})\rangle \langle \psi^{*}_{\kappa^\prime} (-{r})|)$ can only be different up to a phase.

Let us calculate
\begin{equation}
    x^2(\alpha) = \frac{1}{2}\operatorname{Tr}(\vec{X} \cdot \vec{\mu})^2 = v_F(\alpha)^2 - \langle \psi^{*}_{\kappa} (-{r}) |\psi^{}_{\kappa^\prime} ({r})\rangle \langle \psi^{*}_{\kappa^\prime} (-{r})|\psi^{}_{\kappa} ({r})\rangle = v_F(\alpha)^2
\end{equation}
In the third equity, we used the fact that the second term is an integral of motion which does not depend on $\mathbf{r}$ and it vanishes at $\mathbf{r} = \mathbf{r}_0 = \frac{1}{3}(\mathbf{a}_1 - \mathbf{a}_2)$ so it is zero everywhere. Also, we have choose the overall phase of $\vec{X}\cdot \vec{\mu}$ such that $x^2$ is real.

More generally, we can construct $M$ with vanishing trace as
\begin{equation}
    M(\mathbf{r}) = 
    \begin{pmatrix} |\psi^{}_{K} ({r})\rangle & |\psi^{}_{-K} ({r})\rangle  \end{pmatrix} R^{-1} \sigma_z R
    \begin{pmatrix} \langle \psi^{*}_{K} (-{r})| \\  \langle \psi^{*}_{-K} (-{r}) \end{pmatrix}
\end{equation}
where $R$ represents any invertible $2\times2$ matrix.
It has the same property of $x = v_F$ and gives us an additional gauge degree of freedom. We choose $R = I_2$ (identity) without further notice. Also, when calculating the $U(1)$ gauge potential of Eq.~\ref{eq:U1gaugepotential}, there is another degree of freedom to flip the overall sign of $\vec{X}$. In numerical calculations we find that $X_3$ is almost real. Therefore, we choose the sign such that $\operatorname{Re}X_3 > 0$. This helps to avoid singularity in the denominator $x + X_3$.

The wavefunction satisfies the translation property that $|\psi^{}_{\kappa} \left({r}+{a}_{1,2}\right)\rangle=e^{i {\kappa} {a}_{1,2}} U_\omega |\psi^{}_{\kappa} ({r})\rangle$ with $U_\omega=\operatorname{diag}\left(1, \omega^*\right)$. Therefore, the constant of motion satisfies $\vec{X}\left({r}+{a}_{1,2}\right) \cdot \vec{\mu} = U_\omega\vec{X}({r}) \cdot \vec{\mu} U^\dagger_\omega$. Also, because the symmetry property of the wavefunction: $C_{2y}T|\psi_\kappa\rangle = |\psi_{\kappa^\prime}\rangle$, we have $C_{2y}T \vec{X}\cdot\vec{\mu}C_{2y}T = - \vec{X}\cdot\vec{\mu} $.
It is then straightforward to see that $\mathcal{D}^\prime({r} + {a}_{1, 2}) = U_\omega \mathcal{D}^\prime({r}) U_\omega^\dagger = \mathcal{D}^\prime$, because of the fact that $\mathcal{D}^\prime$ is diagonal. This is indeed the translation property we expect for effective magnetic field with zero total flux in one unit cell. Also, one can check that $\mathcal{D}^\prime$ is odd under $C_{2y} T$, which leads to the property of the effective magnetic field that $B^\prime(-x, y) = -B^\prime(x, y) $.

\section{Momentum space spread and stability of the magic angles} \label{subapp_momlock}

\begin{figure}
\centering
\includegraphics[width=0.75\columnwidth]{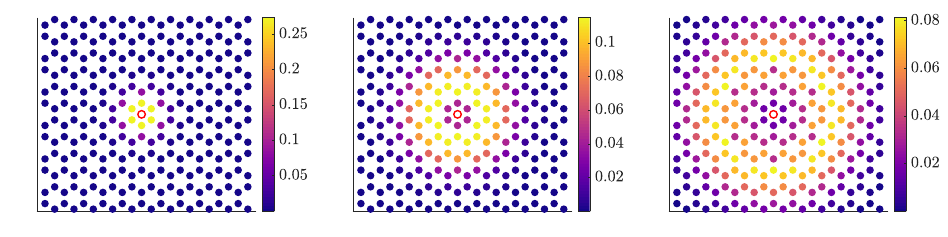}
\caption{Wavefunction at the $\gamma$ point of the Moire BZ on the momentum space lattice for the first three magic angles.}
\label{fig:app:k_lattice_wf}
\end{figure}

Fig.~\ref{fig:app:k_lattice_wf} shows the wavefunction at the $\gamma$ point of the moiré Brillouin zone on the momentum-space lattice for the first three magic angles. It is evident that the momentum distribution of the wavefunction is more spread out for higher magic angles. This broader distribution correlates with the stability of magic angles against perturbations shown in Fig.~\ref{fig_sus_exp}. In the following, we present a perturbative argument to explain this behavior.

The magic angles are determined by the eigenvalues of the operator $\Lambda$, which is expressed as
\begin{equation}
    \Lambda = \hat{D}^{-1} \hat{T}_+ \hat{D}^{-1} \hat{T}_-, \quad [\hat{D}]_{k,k^\prime} = \delta_{k,k^\prime} \frac{1}{k}, \quad [\hat{T}_+]_{k,k^\prime} = \omega^j \delta_{k, k^\prime - q_j}, \quad \hat{T}_- = \hat{T}_+^\dagger.
\end{equation}
Here, $\Lambda$ is a non-Hermitian operator with left and right eigenvectors that satisfy the relations
\begin{equation}
    \Lambda |\psi_{\alpha} \rangle = \frac{1}{\alpha^2} |\psi_\alpha \rangle, \quad \langle \tilde{\psi}_\alpha | \Lambda = \langle \tilde{\psi}_\alpha | \frac{1}{\alpha^2}, \quad \langle \tilde{\psi}_\alpha | \psi_{\alpha^\prime} \rangle = \delta_{\alpha, \alpha^\prime}.
\end{equation}
The magic angles are the $\alpha$ obtained from this equation, which may be real or imaginary. Next, we introduce a perturbation of the form
\begin{equation}
    \hat{T}_\pm \rightarrow \hat{T}_\pm + \delta \hat{T}_\pm,
\end{equation}
where $\delta \hat{T}_\pm$ couples momenta $k$ to $k + Q$, and $Q$ represents a higher momentum, i.e., $|Q| \geq |q_j|$. The first-order correction to the eigenvalues due to this perturbation is given by
\begin{equation}
    \langle \tilde{\psi}_{\alpha} | \hat{D}^{-1} \hat{T}_+ \hat{D}^{-1} \delta\hat{T}_- | \psi_\alpha \rangle + \langle \tilde{\psi}_{\alpha} | \hat{D}^{-1} \delta\hat{T}_+ \hat{D}^{-1} \hat{T}_- | \psi_\alpha \rangle, \label{eq:disorder-first-order-pt}
\end{equation}
which corresponds to a scattering processes in the momentum space. The value of this expression depends sensitively on the spread of the wavefunction $|\psi_\alpha\rangle$.
More precisely, for a wavefunction of the form $|\psi_\alpha \rangle = \sum_{ G } v_G^\alpha | G \rangle$ or $\langle \psi_\alpha | = \sum_{ G } w_G^\alpha \langle G |$, the correction has terms like
\begin{equation}
    \sum\limits_{G, G^\prime} w_{G^\prime} v_G \dfrac{1}{G^\prime} \left(\sum\limits_j  \dfrac{\omega^j}{G^\prime + q_j} \right) \langle G^\prime + q_j | \delta T_- | G \rangle
\end{equation}
which clearly vanish if $\delta T_-$ connects $G$ to $G+Q$ where $|Q| \gg k_\xi$ where $k_\xi$ is the extent of the wavefunction in momentum space.
As we show in Fig.\ref{fig:app:k_lattice_wf}, the first magic angle, $\alpha = 0.586$, has the most localized wavefunction with the smallest $k_\xi \approx |q_j|$. Consequently, the first-order correction vanishes for almost all higher harmonics, consistent with Fig.~\ref{fig_sus_exp} in the main text. The same reasoning can further be applied to the second-order correction, which would scale as $1/Q^2$. The suppression arises from the two powers of the propagator $\hat{D}^{-1}$ in Eq.~\ref{eq:disorder-first-order-pt}. 

In contrast, the second and higher magic angles have wavefunctions that are more spread out in momentum space. There are many contributions to Eq.~\ref{eq:disorder-first-order-pt} and as a result, these wavefunctions are more susceptible to the effects of disorder from this perturbative standpoint.

\section{Details of disorder calculation} 
\label{app_detailsnumerics}

In this Appendix, we provide details on our numerical simulation of disorder in the chiral TBG Hamiltonian. 
We use the method introduced in Ref.~\cite{becker2022mathematics}, which we briefly summarize in Sec.~\ref{subapp_numericalmethod} for completeness. We then analyze the results of the main text showing the robustness of the first magic angle using localization in momentum space (Sec.~\ref{subapp_momlock}).
Finally, we provide additional results for disorders defined on length scale smaller than the moir\'e lattice constant (Sec.~\ref{subapp_numericalmethod}).

\subsection{Numerical method} \label{subapp_numericalmethod}

\begin{figure}
\centering
\includegraphics[width=\columnwidth]{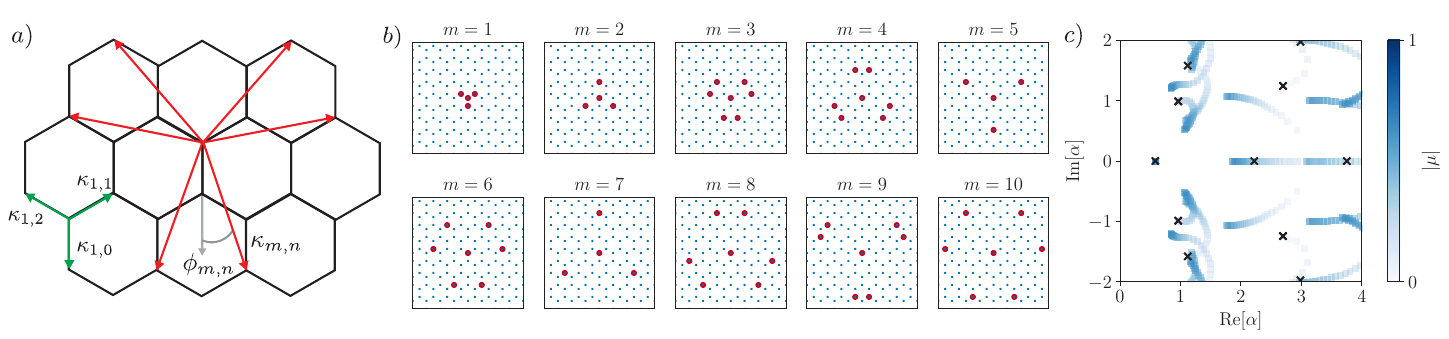}
\caption{a) Definition of the angle $\phi_{m, n}$ as the polar angle of $\kappa_{m, n}$, measured from the negative $y$-axis, represented for the the $ m = 3$ harmonics (red arrows). We also show the lowest harmonics corresponding to the tunneling in the Bistritzer-MacDonald model in green. 
b) Definition of the $\kappa_{m, n}$ vectors, which start from the origin and point to the red dots. c) The flow of magic angles on the complex plane as a function of increasing disorder strength $|\mu|$ (color bar) while $\hat{\mu}$ is fixed. We put disorder in the first $6$ channels and use a truncation $R_k = 20 \sqrt{3} \kappa$.}
\label{fig:app:harmonics}
\end{figure}

We begin by describing the chiral model without disorder. The Hamiltonian can be written as
\begin{equation}
    \mathcal{H}_{\rm D} = \begin{pmatrix} 0 & \mathcal{D} \\ \mathcal{D}^\dag & 0 \end{pmatrix}, \quad {\rm where} \quad \mathcal{D} = \begin{pmatrix}
        \hat{D} & \alpha \hat{T}_- \\ \alpha \hat{T}_+ & \hat{D}
    \end{pmatrix}
\end{equation}
and $\hat{T}_\pm = \hat{T}(\pm z)$ is the inter-layer hopping written in terms of the complex variable $z = x+i y$.
The tunneling potential $\hat{T}$ is related to the gauge field $\mathcal{A}_z$ defined in Eq.~\ref{eq_chirltbg} of the main text by
\begin{equation}
\hat{T}(z) = A_z^{(1)} + i A_z^{(2)} = \sum_{n = 0, 1, 2}\omega^{n} \exp(i\kappa_n\cdot r)
\end{equation}
with $\omega = e^{ i 2\pi/3}$, and where $\kappa_{n} \equiv \kappa_{1,n}$ are the three smallest vectors connecting the two corners of the moir\'e Brillouin zone, represented with green arrows in Fig.~\ref{fig:app:harmonics}a.
We consider the effects of disorder in the inter-layer tunneling potential $\hat{T}_{\pm}$ through the addition of a perturbation that we decompose in harmonics labeled by $m$
\begin{equation} \label{appeq_perturbedhigherharmonics}
\hat{T} \rightarrow \hat{T} + \sum\limits_{m} \mu_m \hat{T}_m
\end{equation}
where $\{\mu_m \}$ is a random vector of small norm $|\mu| = [ \sum_m \mu_m^2]^{1/2} \ll 1$. 
The wave-vectors corresponding to the smallest ten harmonics ($m\leq 10$) are depicted in Fig.~\ref{fig:app:harmonics}b. These channels were used in Fig.~\ref{fig_sus_exp} of the main text as well. Notice that, for each $m$, there are several wave-vectors of equal norm related by $C_3$ and/or $C_{2y}$ symmetry that contribute to $\hat{T}_m$. We denote them as $\kappa_{m, n}$, with $n$ an auxiliary index taking either three or more distinct values depending on the number of nonequivalent wave-vectors in the $m$-th harmonic (compare for instance $m=2$ and $m=3$ in Fig.~\ref{fig:app:harmonics}b). 
In the main text, we considered perturbations $\hat{T}_m$ that preserve the $C_3$ and $C_{2y}$ symmetry of the chiral Hamiltonian for TBG (Eq.~\ref{eq_chirltbg}), which constrains the functional form of these inter-layer tunneling terms to
\begin{equation}
\label{eq_kappa_phi}
\hat{T}_m = a_{m,z}^{(1)} + i a_{m,z}^{(2)} = \sum_{n} e^{i (\kappa_{m,n} \cdot r +\phi_{m, n})} . 
\end{equation}
As illustrated in Fig.~\ref{fig:app:harmonics}a and b, these terms can be interpreted as long-range tunnelings in momentum space, with an amplitude equal to $\mu_m$, a direction given by $\kappa_{m, n}$, and a phase $\phi_{m, n}$. 
Amongst all gauge equivalent choices for the $\phi_{m, n}$ phases leading to $C_3$ and $C_{2y}$ symmetric $\hat{T}_m$, we choose the one where $\phi_{m, n}$ is the polar angle of $\kappa_{m, n}$ measured from the negative $y$-axis, as illustrated in Fig.~\ref{fig:app:harmonics}a. 
The main advantage of this prescription is that it equally applies to harmonics containing three and six wave-vectors (Fig.~\ref{fig:app:harmonics}b).

The problem of finding the magic angle corresponds to finding vectors $|\psi\rangle$ and $|\chi\rangle$ such that
\begin{equation}
\begin{pmatrix}
\hat{D} & \alpha \hat{T}_- \\ \alpha \hat{T}_+ & \hat{D}
\end{pmatrix} \begin{pmatrix}
|\psi\rangle \\ |\chi\rangle
\end{pmatrix} = 0 \quad \Rightarrow \quad \Lambda |\psi\rangle = \dfrac{1}{\alpha^2} |\psi \rangle , \quad \Lambda =  \hat{D}^{-1} \hat{T}_+ \hat{D}^{-1} \hat{T}_- ,
\end{equation}
which itself reduces to finding the eigenvalues $\{ \lambda_i \}$ of the (non-Hermitian) operator $\Lambda$.
The magic angles $\{\alpha_i\}$ are then calculated as $\alpha_i = 1/\sqrt{\lambda_i}$.
The operators $\hat{D}$ and $\hat{T}_{\pm}$ can be represented as sparse matrices on a momentum space grid and are, in practice, truncated by removing all momenta with norm higher than a certain cutoff $R_k$.

As an illustration of the method, we introduce disorder in the first $m_c = 6$ channels using $\vec{\mu} = |\mu| \hat{\mu}$, varying the overall strength of the disorder $|\mu|$ for a fixed random vector normalized to unity $\hat{\mu}$. 
In Fig.~\ref{fig:app:harmonics}c, we show the evolution of the magic angles$\{\alpha_i\}$ on the complex plane as a function of increasing disorder strength $|\mu|$. We see that only the first magic angle is robust against the perturbation, while all the other magic angles are dramatically changed as disorder strength increases.

\subsection{Disorder $C_3$-symmetric on average}

\begin{figure}
\centering
\includegraphics[width=0.7\columnwidth]{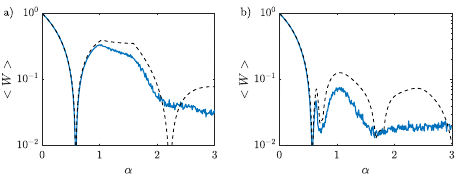}
\caption{Average bandwidth of the lowest band $<W>$ versus $\alpha$ for a) the chiral model ($w_0 = 0$) and b) away from chiral limit $w_0 = 0.8 w$ with the presence of $C_3$ breaking disorder. The black dashed line denotes the quantity without disorder. Disorder strength $\delta \mu = 0.1$, $R_k = 5\sqrt{3}|\kappa|$ and $50$ disorder realizations are averaged over. }
\label{fig:app:C3breaking}
\end{figure}

The disorder we introduced in Sec.~\ref{subapp_numericalmethod} is $C_3$ symmetric by definition. The robustness of the first magic angle against these disorder has been elaborate in Sec.~\ref{sec_numerics} in the main text. Here we investigate disorder that breaks $C_3$ symmetry, and provide numerical evidence that the first magic angle enjoys the same protection while higher magic angles do not. To implement $C_3$ breaking disorder, we simply modify Eq.~\eqref{appeq_perturbedhigherharmonics} that $\delta \hat{T}  = \sum_{mn}\mu_{mn}\hat{T}_{mn} = $ with $\hat{T}_{mn} = e^{i (\kappa_{m,n} \cdot r +\phi_{m, n})}$ and $\{\mu_{mn}\}$ is a random vector of small norm $\delta \mu \equiv|\mu| = [ \sum_{mn} \mu_{mn}^2]^{1/2} \ll 1$ as before. With broken $C_3$ symmetry, the positions of the Dirac cones may be shifted so that it is difficult to characterize the Fermi velocity. Instead, we calculate the average bandwidth of the lowest band, as shown in Fig.~\ref{fig:app:C3breaking}. The calculation conveys a clear message that only the first magic angle is robust against higher harmonic disorder, even to those that breaks $C_3$ symmetry.

\subsection{Small momentum perturbation} \label{subapp_numericalmethod}

The inter-layer tunneling decomposed into higher harmonic $m\geq 1$ of Eq.~\ref{appeq_perturbedhigherharmonics} can only describe disorder with a correlation length smaller than moir\'e unit cell. In this Appendix, we explore the opposite regime where the disorder correlation length is larger than the moir\'e unit cell. Physically, twist angle variations across the sample belong to this class of perturbations.

\begin{figure}
\centering
\includegraphics[width=0.55\columnwidth]{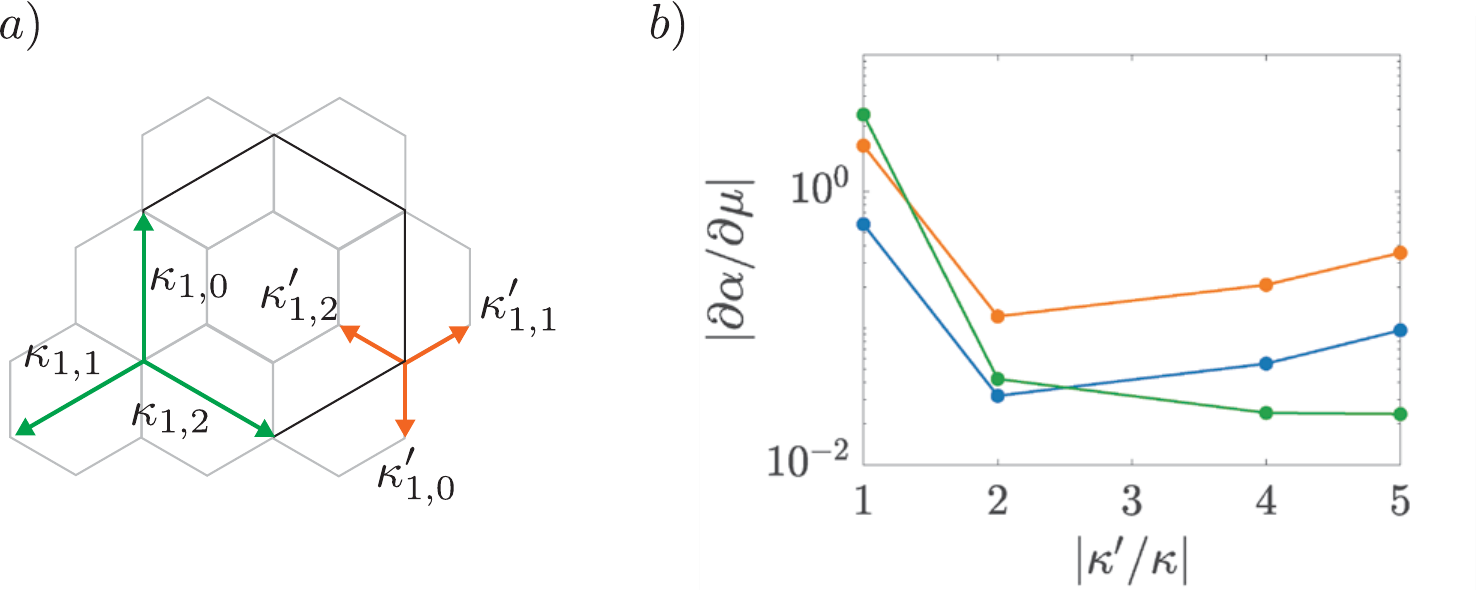}
\caption{a) Illustration of the rescaled honeycomb lattice for $M=-2$ (grey) lines. The rescaled lattice is generated by the basis vectors $\kappa_{1, n}^{\prime}$ with $|\kappa^\prime_{1,n}| = |\kappa|/2$, and contains the wave-vectors $|\kappa_{1,n}|$ appearing in the Bistritzer-Macdonald model. The original moir\'e Brillouin zone is drawn in black. b) Variation $|\partial \alpha / \partial \mu|$ of the first (blue), second (orange), and third (green) magic angle, numerically evaluated using $\delta \mu = 0.02$, as a function of $M = \left| \frac{\kappa}{q_{\rm pert}} \right| = \left| \frac{\kappa}{\kappa_{1,0}^\prime} \right|$. To obtain converged results for large $m$, we used a large truncation radius $R_k = 35 \sqrt{3} |\kappa|$.}
\label{fig:app:lowharm}
\end{figure}

Introducing harmonics that are periodic on an enlarged moir\'e unit cell in real space requires access to wave-vectors that are fractions of the original $\kappa_{1,n}$ appearing in the Bistritzer-MacDonald model (Eq.~\ref{eq_chirltbg}). 
To achieve this numerically, we rescale the momentum-space lattice of Fig.~\ref{fig:app:harmonics}b by a factor $r=1/M$, with $M=3p+1$ and $p\in \mathbb{Z}$. This choice ensures that the reduced lattice still contains the $\kappa_{1,n}$. 
By convention, rescaling by a negative value $r$ corresponds to a rescaling by $|r|>0$ followed by a $180^\circ$ degree rotation, as illustrated in Fig.~\ref{fig:app:lowharm}a for the case $r=-1/2$ (corresponding to $M=-2$, $p=-1$). 
The inter-layer harmonics $\hat{T}_m'$ on the rescaled lattice with wave-vectors $\kappa_{m, n}^\prime$ can now be used to introduce disorder with correlation length larger than the moir\'e lattice constant. The definitions of $\hat{T}_m'$ and $\kappa_{m, n}^\prime$ are the same as $\hat{T}_m$ and $\kappa_{m, n}$ (see App.~\ref{app_detailsnumerics}), albeit on the rescaled honeycomb lattice. We also highlight that $\kappa_{1,n} = \kappa_{|M|, n}^\prime$, which is also illustrated in Fig.~\ref{fig:app:lowharm}a. 

We reproduce the calculation of Fig.~\ref{fig_sus_exp}a for the fractional wave-vectors $\kappa_{m, n}^\prime$ and consider an inter-layer tunneling of the form 
\begin{equation}
\hat{T}^{\prime} = \hat{T}^{\prime}_{|M|} + \delta \mu \hat{T}_1',
\end{equation}
where $\hat{T}^{\prime}_{|M|}$ is nothing but the terms appearing in the un-perturbed chiral Hamiltonian for TBG (Eq.~\ref{eq_chirltbg}), while $\hat{T}_1'$ is the lowest accessible fractional harmonic on the rescaled lattice. The latter realizes a perturbation of momentum $q_{\rm pert} = |\kappa/M|$.
In Fig.~\ref{fig:app:lowharm}b, we show the derivative of the magic angles $|\partial \alpha  / \partial \mu|$, numerically computed using a small disorder strength $\delta\mu = 0.02$ (see Sec.~\ref{sec_numerics}), for the first three magic angles as a function of $M = \left| \kappa/q_{\rm pert} \right| = \left| \kappa/\kappa_{1,0}^\prime \right|$.
We observe that, unlike the perturbation of larger momentum where only the first magic angle enjoyed protection, all magic angles respond similarly to disorder. 
We now understand this behavior from the momentum space picture presented in the previous section.
There is no special protection of the first magic angle against perturbations of correlation length longer than the moir\'e lattice vector.

\end{document}